\documentclass[conference]{IEEEtran}
\IEEEoverridecommandlockouts
\usepackage{cite}
\usepackage{amsmath,amssymb,amsfonts}
\usepackage{algorithmic}
\usepackage{graphicx}
\usepackage{textcomp}
\usepackage{adjustbox}
\usepackage{subcaption}
\usepackage{listings}
\usepackage{color}
\usepackage[bookmarks=false]{hyperref}
\def\fixme#1{\typeout{FIXED in page \thepage : {#1}}
\bgroup \color{red}{[FIXME: {#1}]} \egroup}

\def\BibTeX{{\rm B\kern-.05em{\sc i\kern-.025em b}\kern-.08em
    T\kern-.1667em\lower.7ex\hbox{E}\kern-.125emX}}
\begin{document}

\title{DeepPicar: A Low-cost Deep Neural Network-based Autonomous Car}

\author{Michael G. Bechtel$^\dagger$, Elise McEllhiney$^\dagger$, Minje Kim$^\star$, Heechul Yun$^\dagger$\\
  $^\dagger$ University of Kansas, USA. \{mbechtel, elisemmc, heechul.yun\}@ku.edu \\
  $^\star$ Indiana University, USA. minje@indiana.edu \\
}

%% \IEEEauthorblockA{\textit{Electrical Engineering and Computer Science} \\
%% \textit{University of Kansas}\\
%% Lawrence, USA \\
%% mbechtel@ku.edu}
%% \and
%% \IEEEauthorblockN{Elise McEllhiney}
%% \IEEEauthorblockA{\textit{Electrical Engineering and Computer Science} \\
%% \textit{University of Kansas}\\
%% Lawrence, USA \\
%% e908m428@ku.edu}
%% \and
%% \IEEEauthorblockN{Minje Kim}
%% \IEEEauthorblockA{\textit{Electrical Engineering and Computer Science} \\
%% \textit{Indiana University}\\
%% Bloomington, USA \\
%% minje@indiana.edu}
%% \and
%% \IEEEauthorblockN{Heechul Yun}
%% \IEEEauthorblockA{\textit{Electrical Engineering and Computer Science} \\
%% \textit{University of Kansas}\\
%% Lawrence, USA \\
%% heechul.yun@ku.edu}
%% }

\maketitle

\begin{abstract}
We present DeepPicar, a low-cost deep neural network based
autonomous car platform. DeepPicar is a small scale
replication of a real self-driving car called DAVE-2 by NVIDIA. DAVE-2
uses a deep convolutional neural network (CNN),
which takes images from a front-facing camera as input and produces
car steering angles as output. DeepPicar uses the same
network architecture---9 layers, 27 million connections and 250K
parameters---and can drive itself in real-time using a web camera and
a Raspberry Pi 3 quad-core platform.
Using DeepPicar, we analyze the Pi 3's computing capabilities to
support end-to-end deep learning based real-time control of autonomous
vehicles.
We also systematically compare other contemporary embedded
computing platforms using the DeepPicar's CNN-based real-time control
workload.

We find that all tested platforms, including the Pi 3, are capable of
supporting the CNN-based real-time control, from 20 Hz up to 100
Hz, depending on hardware platform.
However, we find that shared resource contention remains an
important issue that must be considered in applying CNN models on
shared memory based embedded computing platforms; we observe
up to 11.6X execution time increase in the CNN based control loop due
to shared resource contention. To protect the CNN workload, we also
evaluate state-of-the-art cache partitioning and memory bandwidth
throttling techniques on the Pi 3. We find that cache partitioning is
ineffective, while memory bandwidth throttling is an effective
solution.
\end{abstract}


%-------------------------------------------------------------------------
\section{Introduction} \label{sec:intro}

% broad context:
% - advance in ai sparked interests in the robotics application, such as
%   self-driving cars.
% - in particular, deep neural network models are increasingly used
%   for perception and control of a vehicle. say. AI workloads.
%
%
Autonomous cars have been a topic of increasing interest in recent
years as many companies are actively developing related hardware
and software technologies toward fully autonomous driving capability with
no human intervention. Deep neural networks (DNNs) have been
successfully applied in various perception and control tasks in
recent years.  They are important workloads for autonomous vehicles
as well. For example, Tesla Model S was known to use a specialized
chip (MobileEye EyeQ), which used a vision-based real-time obstacle
detection system based on a DNN. More recently, researchers 
are investigating DNN based end-to-end real-time control for
robotics applications~\cite{Bojarski2016,Levine2016}. It is expected
that more DNN based artificial intelligence (AI) workloads may be
used in future autonomous vehicles.

% big problem
Executing these AI workloads on an embedded computing platform 
poses several additional challenges. First, many AI workloads,
especially those in vehicles, are computationally demanding and have
strict real-time requirements.
For example, computing latency in a vision-based object
detection task may be directly linked to the safety of the vehicle. This
requires a high computing capacity as well as the means to guaranteeing
the timings. On the other hand, the computing hardware platform must
also satisfy cost, size, weight, and power constraints, which require a
highly efficient computing platform. These two conflicting
requirements complicate the platform selection process as observed 
in~\cite{Otterness2017}.

%% For example, while today's self-driving car
%% prototype equip more \$100,000
%% of computers and sensors~\cite{juliussen2014emerging}, a study
%% found that aveage consumers are willing to pay much less amount of
%% extra cost for a self-driving capability~\cite{Daziano2017}.
%% https://qz.com/924212/what-it-really-costs-to-turn-a-car-into-a-self-driving-vehicle/

% related work and remaining problems
To understand what kind of computing hardware is needed for AI
workloads, we need a testbed and realistic workloads. While using a
real car-based testbed would be most ideal, it is not only highly
expensive, but also poses serious safety concerns that hinder
development and exploration. Therefore, there is a need for safer and
less costly testbeds.

%% There are already several relatively inexpensive RC-car
%% based testbeds, such as MIT's 
%% RaceCar~\cite{shin2017project} and UPenn's F1$/$10 racecar~\cite{upennf1tenth}.
%% However, these RC-car testbeds still cost more than \$3,000, requiring
%% considerable investment.

%% % our goals
%% Instead, we want to build a low cost testbed that still employs the
%% state-of-the art AI technologies. Specifically, we focus on an end-to-end
%% deep learning based real-time control system,
%% which was developed for a real self-driving car, NVIDIA
%% DAVE-2~\cite{Bojarski2016}, and uses the same methodology on a
%% smaller and less costly setup. In developing the testbed, our
%% goals are (1) to analyze real-time issues in DNN based end-to-end
%% control; and (2) to evaluate real-time performance of contemporary embedded
%% platforms for such a workload.

% DeepPicar introduction
In this paper, we present DeepPicar, a low-cost autonomous car
testbed for research. From a hardware perspective,
DeepPicar is comprised of a Raspberry Pi 3 Model B quad-core
computer, a web camera and a small RC car, all of which are affordable
components (less than \$100 in total).
The DeepPicar, however, employs a state-of-the-art AI
technology, namely end-to-end deep learning based real-time control,
which utilizes a deep convolutional neural network (CNN).
The CNN receives an image frame from a single forward
looking camera as input and generates a predicted steering angle
value as output at each control period in \emph{real-time}.
The CNN has 9 layers, about 27 million connections
and 250 thousand parameters (weights).
DeepPicar's CNN architecture is identical to that of NVIDIA's
real-sized self-driving car, called DAVE-2~\cite{Bojarski2016}, which
drove on public roads without human driver's intervention while only
using the CNN.
Using DeepPicar, we systematically analyze its real-time
capabilities in the context of end-to-end deep-learning based real-time
control, especially on real-time \emph{inferencing} of the CNN.
We also evaluate other, more powerful, embedded computing
platforms to better understand achievable real-time performance of
DeepPicar's CNN based control system and the performance impact of
computing hardware architectures.
From the systematic study, we want to answer the following questions:
(1) How well does the CNN based real-time inferencing task perform on
contemporary embedded multicore computing platforms? (2) How
susceptible is the CNN inferencing task to contention in shared
hardware resources (e.g., cache and DRAM) when multiple tasks/models
are consolidated? (3) Are the existing state-of-the-art shared resource
isolation techniques effective in protecting real-time performance of
the CNN inferencing task?

% Findings:
%
% - we find DNN processing is an ideal real-time workload from the
%   perspective of timing predictability as the amount of computation
%   needed is fixed at design time and doesn't change over input.
%
% - real-time processing of non-trivial DNN inferencing is feasible
%   on today's embedded multicore platform, even as inexpensive as
%   raspberry pi3.
%
% - all tested platforms have enough computing capacity to consolidate 
%   multiple DeepPicar's CNN workloads or other compute intensive tasks
%   simultaneously.
%
% - however, when consolidating multiple workloads shared resource
%   contention is an important issue. we carefully analyzed
%   sensitivity of DeepPicar's workload with respect to shared
%   hardware resources.
%
% - in particular, we find DNN processing is highly sensitive to
%   memory performance. performance severely affected when memory
%   bandwidth is contented.
%
% - however, we find that dnn processing is not sensitive to cache
%   space. we find applying cache partitioning does not help improve
%   isolation of DNN workload when memory bandwidth is contended.
Our main observations are as follows.
First, we find that real-time processing of the CNN inferencing is
feasible on contemporary embedded computing platforms, even one   
as inexpensive as the Raspberry Pi 3. Second, while consolidating
additional CNN models and tasks on a single multicore platform is
feasible, the impact of shared resource contention can be
catastrophic---we observe up to 11.6X slowdown even when the CNN model
was given a dedicated core.
Third, our evaluation of existing cache
partitioning and memory bandwidth throttling
techniques~\cite{yun2014rtas,Yun2013} shows that cache partitioning is
not effective in protecting the CNN inferencing task while memory
bandwidth throttling is quite effective.

This paper makes the following {\bf contributions}:
\begin{itemize}
  \item We present DeepPicar, a low-cost autonomous car testbed, which
    employs a state-of-the-art CNN based end-to-end
    real-time control~\footnote{The source code, datasets, and build
      instructions of DeepPicar can be found at:
      \url{https://github.com/mbechtel2/DeepPicar-v2}}.
  \item We provide extensive empirical performance evaluation results
    of the CNN inferencing workload on a number of contemporary
    embedded computing platforms.
  \item We apply the state-of-the-art shared resource isolation
    techniques and evaluate their effectiveness in protecting the
    inferencing workload in consolidated settings.
\end{itemize}

The remainder of the paper is organized as follows.
Section~\ref{sec:background} provides a background on the application
of neural networks in autonomous driving.
Section~\ref{sec:overview} gives an overview of the DeepPicar
testbed. Section~\ref{sec:evaluation} presents extensive real-time
performance evaluation results we collected on the testbed.
Section~\ref{sec:comparison} offers a comparison between  
the Raspberry Pi 3 and other embedded computing platforms.
%% We discuss more detailed training methods and system-level issues
%% in~\ref{sec:discussion}. 
We review related work in
Section~\ref{sec:related} and conclude in
Section~\ref{sec:conclusion}.

\section{Background} \label{sec:background}

In this section, we provide background on the application of deep
learning in robotics, particularly autonomous vehicles. 

%% \subsection{Deep Convolutional Neural Network}
%% \fixme{CNN background}

\subsection{End-to-End Deep Learning for Autonomous Vehicles}

%% - explosion of AI
%% - in particular, application of DNN in perception and control of robotics systems.
%% - end-to-end control is a promising technique.
%%   levine's publications?
%% - examples: nvidia's DAVE-II prototype, forest navigating drone
%% challenge problem: computing at low cost?

\begin{figure}[h]
  \centering
  \includegraphics[width=.5\textwidth]{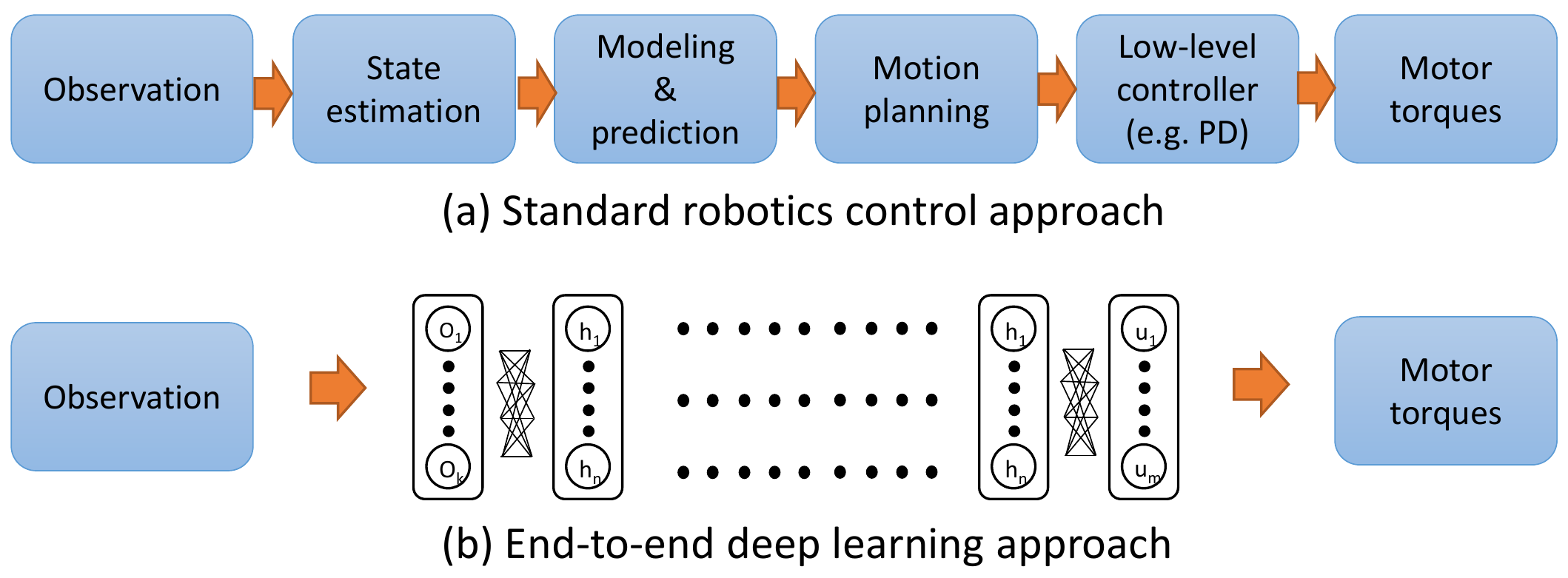}
  \caption{Standard robotics control vs. DNN based end-to-end
    control. Adopted from ~\cite{Levine2017cs294}.}
  \label{fig:end-to-end-control}
\end{figure}

To solve the problem of autonomous driving, a standard approach has
been decomposing the problem into multiple sub-problems,
such as lane marking detection, path planning, and low-level
control, which together form a processing pipeline~\cite{Bojarski2016}.
Recently, researchers have begun exploring another approach that dramatically
simplifies the standard control pipeline by applying deep neural
networks to directly produce control outputs from sensor
inputs~\cite{Levine2016}. Figure~\ref{fig:end-to-end-control}
shows the differences between the two approaches.

The use of neural networks for end-to-end control of autonomous
cars was first demonstrated in the late 1980s~\cite{Pomerleau1989},
using a small 3-layer fully connected neural network; and subsequently
in a DARPA Autonomous Vehicle (DAVE) project in early
2000s~\cite{LeCun:04}, using a 6 layer convolutional neural network
(CNN); and most recently in NVIDIA's DAVE-2
project~\cite{Bojarski2016}, using a 9 layer CNN. In all of these projects,
the neural network models take raw image pixels as input and directly
produce steering control commands, bypassing all intermediary steps and
hand-written rules used in the conventional robotics control approach.  
NVIDIA's latest effort reports that their trained CNN
autonomously controls their modified cars on public roads without human
intervention~\cite{Bojarski2016}.

Using deep neural networks involves two distinct
phases~\cite{NVIDIA2015}. The first
phase is \emph{training}, during which the weights of the network are
incrementally updated by backpropagating errors it sees from the
training examples. Once the network is trained---i.e., the weights of
the network minimize errors in the training examples---the next phase
is \emph{inferencing}, during which unseen data is fed to the network
as input to produce predicted output (e.g., predicted image
classification). In general, the training phase is more computationally
intensive and requires high throughput, which is generally not
available on embedded platforms. The inferencing phase, on the
other hand, is relatively less computationally intensive and latency becomes
as important, if not moreso, as computational throughput, because many
use cases have strict real-time requirements.

%% (e.g., search query latency)

%% \cite{Levine2016}: ``In this paper, we aim to answer
%% the following question: does training the perception and control
%% systems jointly end-toend 
%% provide better performance than training each component separately?''

%% \cite{Bojarski2016} nvidia paper
%% ``We trained a convolutional neural network (CNN) to map raw pixels from
%% a sin- gle front-facing camera directly to steering commands.''

%% ``Compared to explicit decomposition of the problem, such as lane
%% marking detec- tion, path planning, and control, our end-to-end system
%% optimizes all processing steps simultaneously. We''

%% UPenn's f1/10 BOM: $3,628.37
%% http://f1tenth.org/
%% http://selfdrivingcars.mit.edu/
%% http://fast.scripts.mit.edu/racecar/
%% https://github.com/mit-racecar
%% https://mit-racecar.github.io/

\subsection{Embedded Computing Platforms for Real-Time Inferencing}
%% \fixme{we can consider moving this section to a dedicated related work
%%   section at the end.}
Real-time embedded systems, such as an autonomous vehicle, present
unique challenges for deep learning, as the computing platforms of such
systems must satisfy two often conflicting goals:
%% Recent successes in AI, including NVIDIA's DAVE-2 showing, are due
%% in large part to the increased computing performance,
%% which afforded researchers to train and use ever deeper neural networks with
%% high accuracy.
%% For practical applications, the computer platform in an
%% autonomous vehicle must satisfy two often conflicting goals:
(1) The platform must provide 
enough computing capacity for real-time processing of computationally
expensive AI workloads (deep neural networks); and
(2) The platform must also satisfy various
constraints such as cost, size, weight, and power consumption limits~\cite{Otterness2017}.

Accelerating AI workloads, especially inferencing
operations, has received a lot of attention from academia and industry
in recent years as applications of deep learning are broadening to include
areas of real-time embedded systems such as autonomous vehicles. These
efforts include the development of various heterogeneous architecture-based 
system-on-a-chip (SoCs) that may include multiple cores, GPU,
DSP, FPGA, and neural network optimized ASIC hardware~\cite{Jouppi2017}.
Consolidating multiple tasks on SoCs with a lot of shared hardware
resources while guaranteeing real-time performance is also an active
research area, which is orthogonal to improving raw
performance. Consolidation is necessary for efficiency, but unmanaged 
interference can nullify the benefits of consolidation~\cite{Kim2016}.
%% For these reasons, finding a good computing platform is a
%% non-trivial task, one that requires a deep understanding of the
%% workloads and the hardware platform being utilized.

The \emph{primary objectives of this study} are (1) to understand the
necessary computing performance to realize deep neural network based
robotic systems, (2) to understand the characteristics of the
computing platform to support such workloads, and (3) to evaluate the
significance of contention in shared hardware resources and existing
mitigation techniques to address the contention problem.

To achieve these goals, we implement a low-cost autonomous car platform
as a case-study and systematically conduct experiments, which we will 
describe in the subsequent sections.

\section{DeepPicar}\label{sec:overview}

In this section, we provide an overview of our DeepPicar platform.
In developing DeepPicar, one of our primary goals is to faithfully
replicate NVIDIA's DAVE-2 system on a smaller scale using a low cost
multicore platform, the Raspberry Pi 3. Because Raspberry Pi 3's
computing performance is much lower than that of the DRIVE
PX~\cite{drivepx} platform used in DAVE-2, we are interested in if,
and how, we can process 
computationally expensive neural network operations in
real-time. Specifically, inferencing (forward pass processing)
operations must be completed within each control period
duration---e.g., a WCET of 33.3 ms for 30 Hz control 
frequency---locally on the Pi 3 platform, although training of the 
network (back-propagation for weight updates) can be done offline and 
remotely using a desktop computer.

\begin{figure}[h]
  \centering
  \includegraphics[width=.4\textwidth]{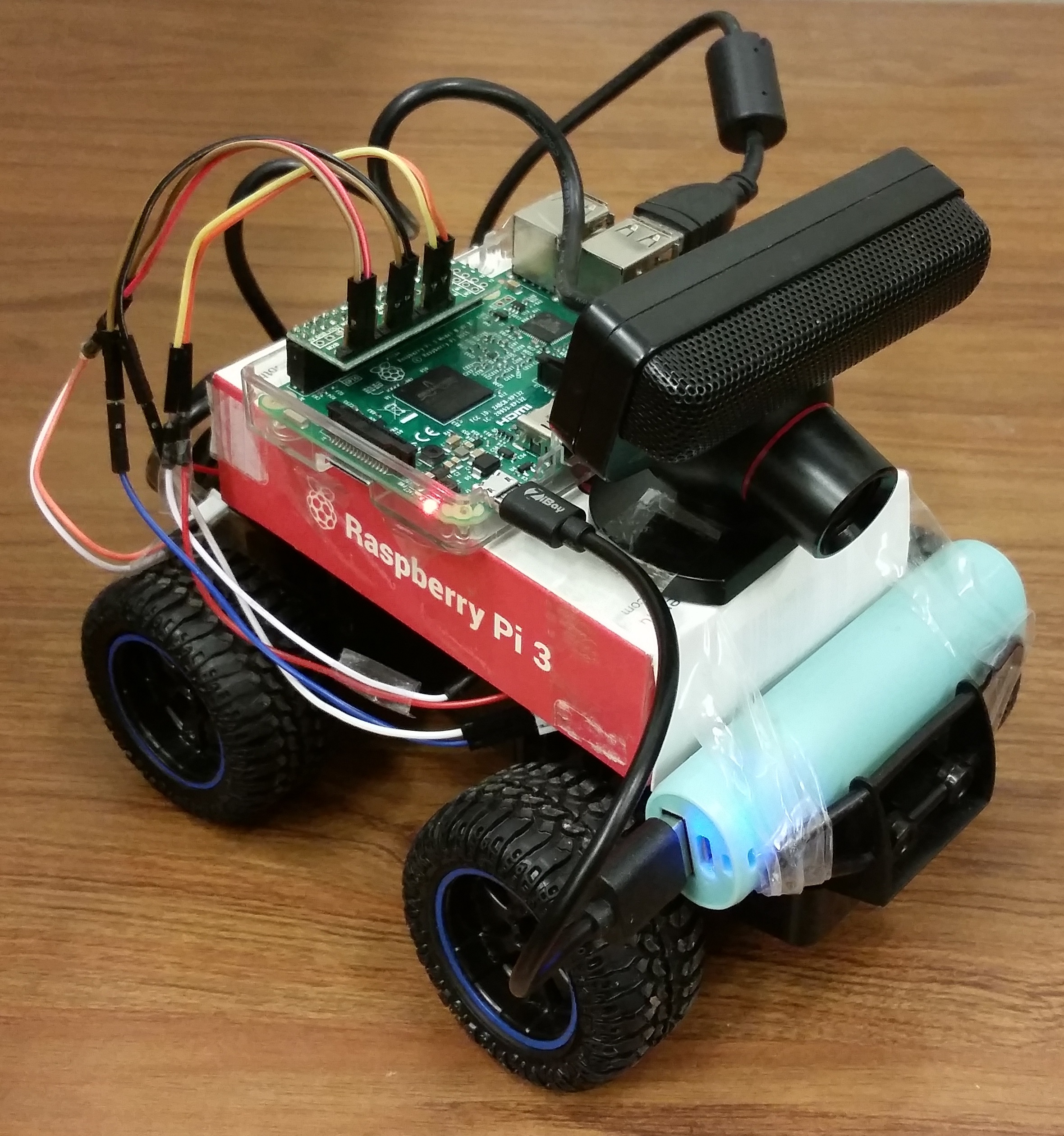}
  \caption{DeepPicar platform.}
  \label{fig:overview}
\end{figure}

\begin{table}[h]
  \centering
  \begin{tabular}{|c|r|}
    \hline
    Item                    & Cost (\$) \\
    \hline
    Raspberry Pi 3 Model B  & 35 \\
    New Bright 1:24 scale RC car       & 10 \\
    Playstation Eye camera  &  7 \\
    Pololu DRV8835 motor hat&  8 \\
    External battery pack \& misc.   & 10 \\
    \hline
    Total                   & 70 \\
    \hline
  \end{tabular}
  \caption{DeepPicar's bill of materials (BOM)}
  \label{tbl:carbom}
\end{table}

Figure~\ref{fig:overview} shows the DeepPicar, which is comprised of a
set of inexpensive components: a Raspberry Pi 3 Single Board Computer
(SBC), a Pololu DRV8835 motor driver, a Playstation Eye webcam, a
battery, and a 1:24 scale RC car. Table~\ref{tbl:carbom} shows the
estimated cost of the system.

\begin{figure}[h]
  \centering
  \includegraphics[width=0.4\textwidth]{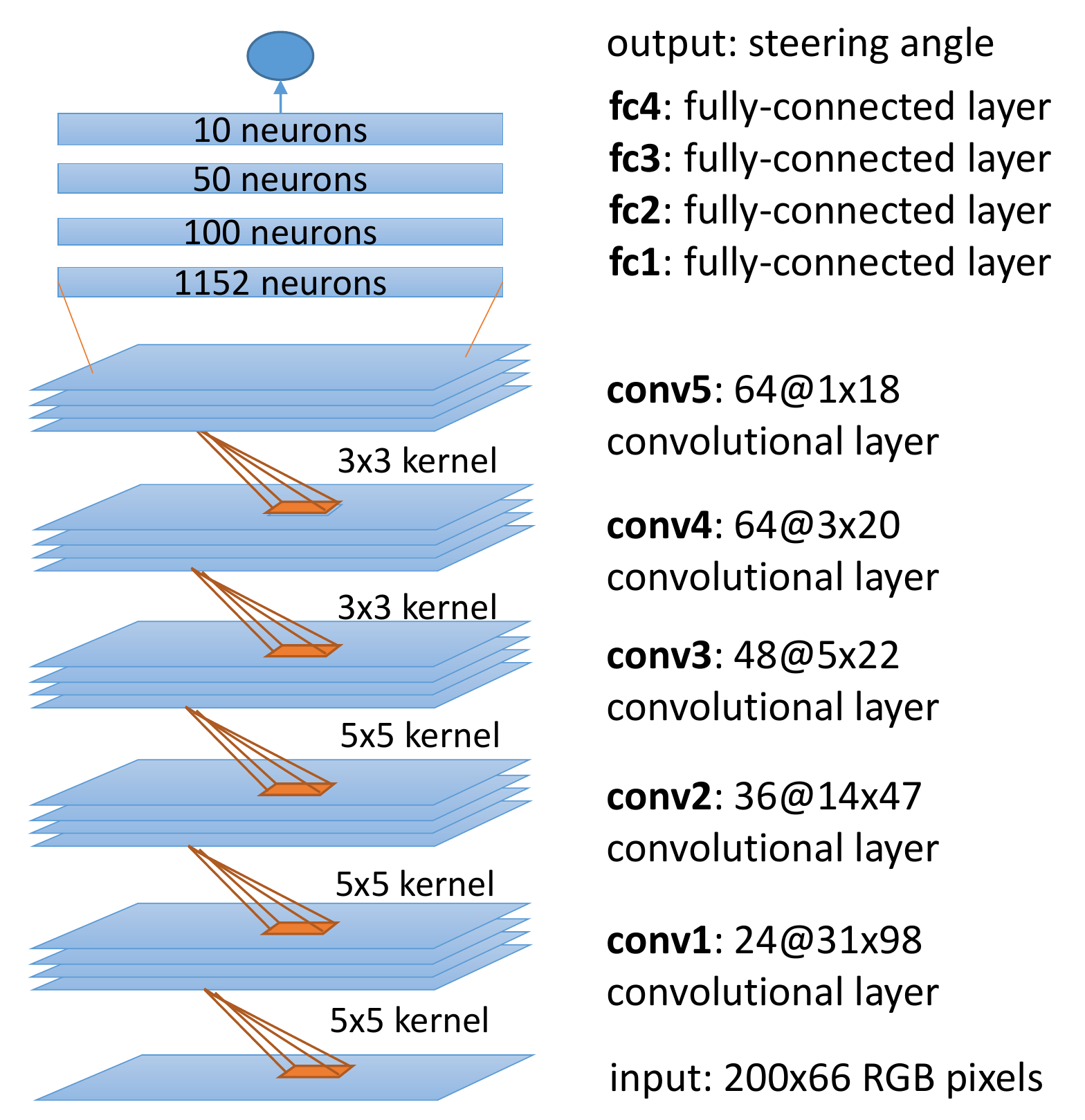}
  \caption{DeepPicar's neural network architecture: 9 layers (5
    convolutional, 4 fully-connected layers), 27 million connections,
    250K parameters. The CNN architecture is identical to the one 
	used in NVIDIA's real self-driving car~\cite{Bojarski2016}.}
  \label{fig:architecture}
\end{figure}

For the neural network architecture, we implement NVIDIA DAVE-2's
convolutional neural network (CNN) using an open-source CNN model in
~\cite{deeptesla}. Note, however, that the CNN model
in~\cite{deeptesla} is considerably larger than NVIDIA's CNN
model as it contains an additional fully-connected layer of
approximately 1.3M additional parameters. We remove the additional
layer to faithfully recreate NVIDIA's original CNN model.
%% The main difference is that we do not utilize a normalization layer, and 
%% instead initialize the weights using a Xavier initialization. 
As in DAVE-2, the CNN takes a raw color image (200x66 RGB pixels)
as input and produces a single steering angle value as output.
Figure~\ref{fig:architecture} shows the network architecture
used in this paper, which is comprised of 9 layers, 250K parameters,
and about 27 million connections as in NVIDIA DAVE-2's architecture.

%% Note, however,
%% that we did not apply the normalization mentioned
%% in~\cite{Bojarski2016}, as it does not include trainable parameters
%% and its computational demand with respect to the overall CNN
%% processing is minimal.

\begin{figure}[h]
  \centering
  \begin{subfigure}{0.4\textwidth}
    \includegraphics[width=\textwidth]{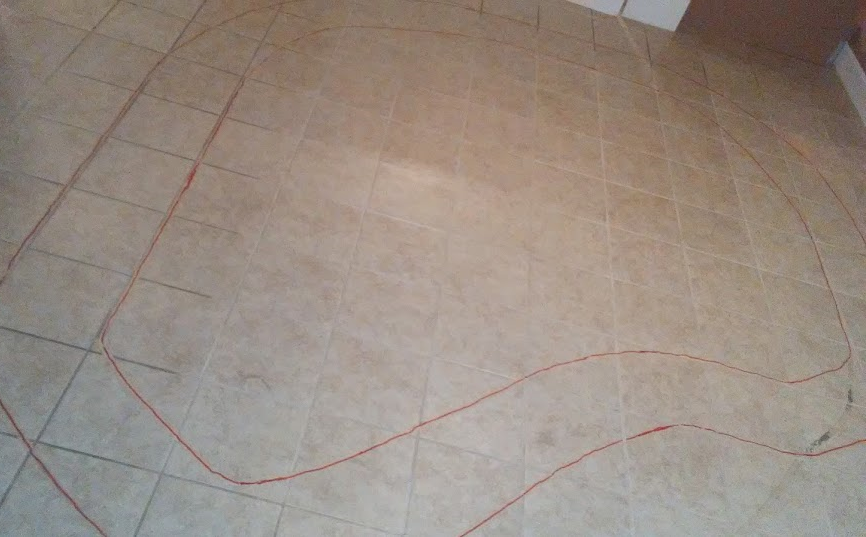}
    \caption{Track 1.}
    \label{fig:track}
  \end{subfigure}
  \hfill
  \begin{subfigure}{0.4\textwidth}
    \includegraphics[width=\textwidth]{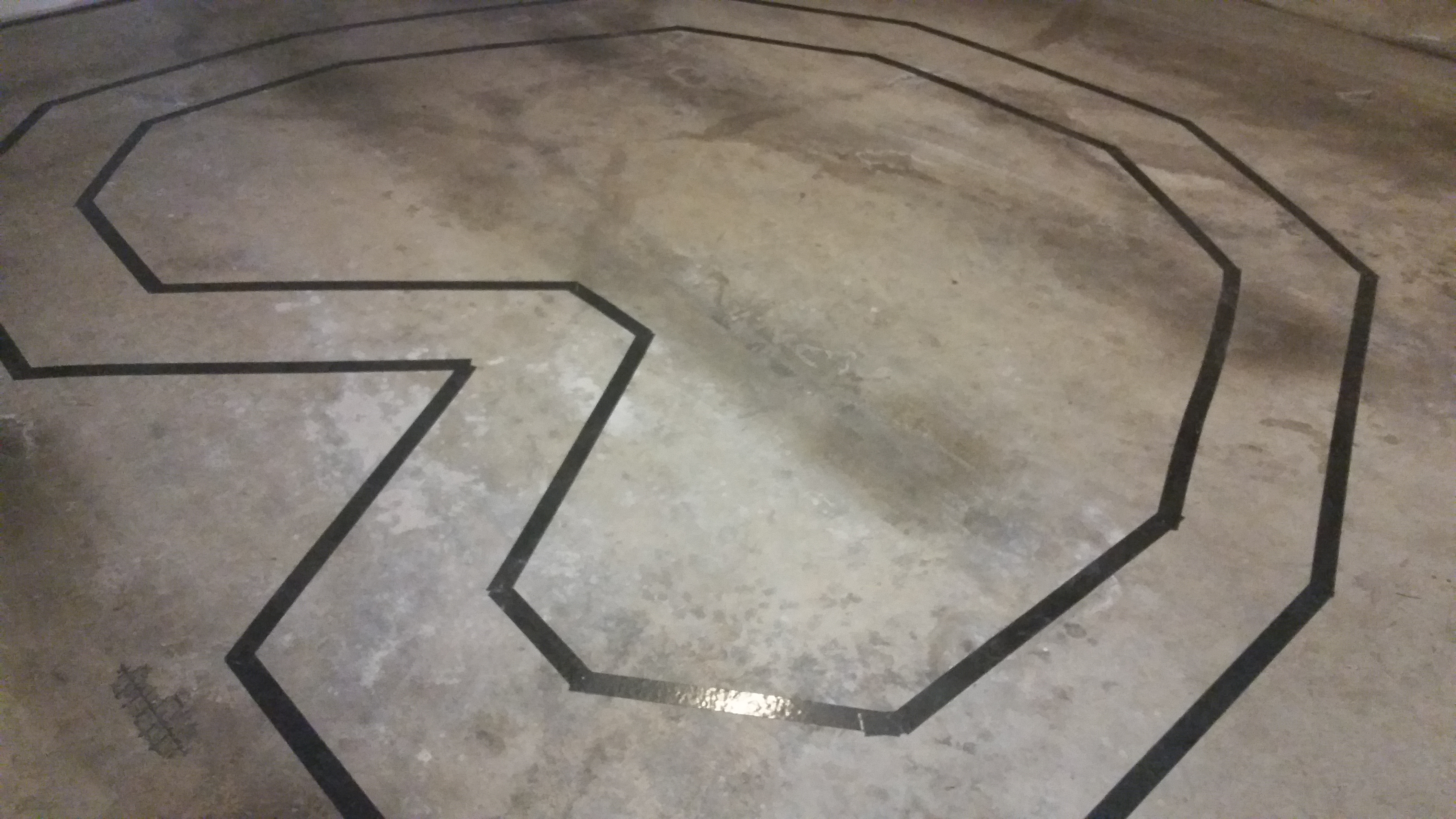}
    \caption{Track 2.}
    \label{fig:track2}
  \end{subfigure}
  \caption{Custom tracks used for training/testing}
  \label{fig:tracks}
\end{figure}

% data collection and training.
To collect the training data, a human pilot manually drives the RC car
on tracks we created (Figure~\ref{fig:tracks}) to record
timestamped videos and contol commands. The stored data is then copied 
to a desktop computer, which is equipped with a NVIDIA Titan Xp GPU, 
where we train the network to accelerate training speed.

%% For comparison, training the network on the Raspbeery Pi 3 takes
%% approximately 4 hours, whereas it takes only about 4 minutes on the
%% desktop computer using the Titan Xp GPU.

\begin{figure}[h]
   \lstset{language=python,
           basicstyle=\ttfamily\small,
           keywordstyle=\color{blue}\ttfamily,
           stringstyle=\color{red}\ttfamily,
           commentstyle=\color{green}\ttfamily
          }  
\begin{lstlisting}[language=python]

while True:
  # 1. read from the forward camera
  frame = camera.read()
  # 2. convert to 200x66 rgb pixels
  frame = preprocess(frame)
  # 3. perform inferencing operation
  angle = DNN_inferencing(frame)
  # 4. motor control                       
  steering_motor_control(angle)
  # 5. wait till next period begins                       
  wait_till_next_period() 
  
\end{lstlisting}
  \caption{Control loop}
  \label{fig:controlloop}
\end{figure}

% inferencing on pi3
Once the network is trained on the desktop computer, the trained model
is copied back to the Raspberry Pi 3. The network is then used
by the car's main controlller, which feeds image frames from the web
camera as input to the network in real-time. At each control period,
the network produced steering angle output is converted into the PWM values
of the car's steering motor. Figure~\ref{fig:controlloop} shows simplified 
pseudo code of the controller's main loop. Among the five steps, the 3rd step, 
network inferencing, is the most computationally intensive and is expected to 
dominate the execution time.

Note that although the steering angle output of the CNN ($angle$) is
a continuous real value, the RC car we used only supports
three discrete angles---left (-30$^{\circ}$), center 
(0$^{\circ}$), and right (+30$^{\circ}$)---as control inputs.
We approximate the network generated real-valued angle to the closest
one of the three angles. Although this may introduce inaccuracy in
control, the observed control performance of the system is respectable,
likely due to the inherently stochastic nature of CNNs.

%% In the future, we plan to use a different (more expensive) RC car
%% platform that can precisely control the car's steering angle.

Other factors that can potentially affect the prediction accuracy of
the CNN, are camera and actuator (motor) control latencies. The camera
latency is defined as the time the camera sensor observes the scene to the
time the computer actually reads the digitized image data. This time
can be noticable depending on the camera used and the data processing
time of the computing platform. Higher camera latency could
negatively affect control performance, because the CNN would analyze
stale scenes. The actuator (motor) control latency is defined as the time
the control output is sent to the steering motor to the time the motor
actually moves to a desired position, which also can take
considerable time. In our platform, the combined latency is measured
to be around 50 ms, which is reasonable.
%% In other words, %% in the recorded video and control data, 
%% a control action of the CNN appears to be applied 50 ms later.
%% Considering that camera's framerate is 30Hz
%% (33.3 ms/frame), this is about two frames after the control action.
%% We experimentally measured the camera
%% latency and found it to be around 50-100 ms.
If this value is too high, control performance may suffer.
Our initial prototype suffered from this problem as the observed latency
was as high as 300 ms, which negatively affected control performance.
For reference, the latency of human perception is known to be as fast
as 13 ms~\cite{ThomasBurger2015}. 
% https://www.pubnub.com/blog/2015-02-09-how-fast-is-realtime-human-perception-and-technology/

Our trained CNN models showed good prediction accuracy, successfully
navigating several different tracks we used for training.
For instance, the DeepPicar could remain on Track 1
(Figure~\ref{fig:track}) for over 10 minutes at a moderate speed (50\%
throttle), at which point we stopped the experiment, and more than one
minute at a higher speed (75\% throttle)~\footnote{Self-driving videos: \url{https://photos.app.goo.gl/q40QFieD5iI9yXU42}
  %% \url{https://photos.app.goo.gl/ce93sU7jPk4ywO8u2}
}. Running at
higher speed is inherently more challenging because the CNN controller
has less time to recover from mistakes (bad predictions).  Also, we
find that the prediction accuracy is significantly affected by the
quality of training data as well as various environmental factors such
as lighting conditions. We plan to investigate more systematic ways
to improve the CNN's prediction accuracies.

We would like to stress, however, that
%% the issues related to the
%% CNN's accuracies have no impact on the \emph{computational 
%% aspects of the system}, and that 
our main focus of this study is not in improving the network accuracy
but in closely replicating the DAVE-2's network architecture and
studying its real-time characteristics, which will be presented in the
subsequent section.

\section{Evaluation}\label{sec:evaluation}

In this section, we experimentally analyze various real-time aspects
of the DeepPicar. This includes
(1) measurement based worst-case execution time (WCET) analysis of
deep learning inferencing,
(2) the effect of using multiple cores in accelerating inferencing,
(3) the effect of co-scheduling multiple deep neural network models,
and
(4) the effect of co-scheduling memory bandwidth intensive co-runners,
and
(5) the effect of shared L2 cache partitioning and memory bandwidth
throttling for guaranteed real-time performance.

\subsection{Setup}
The Raspberry Pi 3 Model B platform used in DeepPicar equips a Broadcom
BCM2837 SoC, which has a quad-core ARM Cortex-A53 cluster,
running at up to 1.2GHz. Each core has private 32K I/D caches
and all cores share a 512KB L2 cache.
The chip also includes Broadcom's Videocore IV GPU, although we do
not use the GPU in our evaluation, due to the lack of sofware support
\footnote{TensorFlow currently only supports NVIDIA's GPUs.}.
For software, we use Ubuntu MATE 16.04 and TensorFlow 1.1.
We disable DVFS (dynamic voltage frequency scaling) and
configure the clock speed of each core statically at the maximum 1.2GHz.
We use the SCHED\_FIFO real-time scheduler to schedule the CNN control
task while using the CFS when executing memory intensive co-runners.

\subsection{Inference Timing for Real-Time Control}
For real-time control of a car (or any robot), the control loop
frequency must be sufficiently high so that the car can quickly
react to the changing environment and its internal states. In general,
control performance improves when the frequency is higher, though
computation time and the type of particular physical system are
factors in determining a proper control loop frequency. While a standard
control system may be comprised of multiple control loops with
differing control frequencies---e.g., an inner control loop for lower-level
PD control, an outer loop for motion planning, etc.---DeepPicar's
control loop is a single layer, as shown earlier in
Figure~\ref{fig:controlloop}, since a single deep neural network
replaces the traditional multi-layer control pipline. (Refer to
Figure~\ref{fig:end-to-end-control} on the differences between the
standard robotics control vs. end-to-end deep learning approach).
This means that the CNN inference operation must be completed
within the inner-most control loop update frequency.

To understand achievable control-loop update frequencies, we
experimentally measured the execution times of DeepPicar's CNN
inference operations.

% 50-200Hz for quadcopters:
% https://robotics.stackexchange.com/questions/231/what-frequency-does-my-quadcopter-output-sense-calculate-output-update-loop-need
% https://quadmeup.com/pid-looptime-why-it-is-not-only-about-frequency/

\begin{figure}[h]
  \centering
  \includegraphics[width=.45\textwidth]{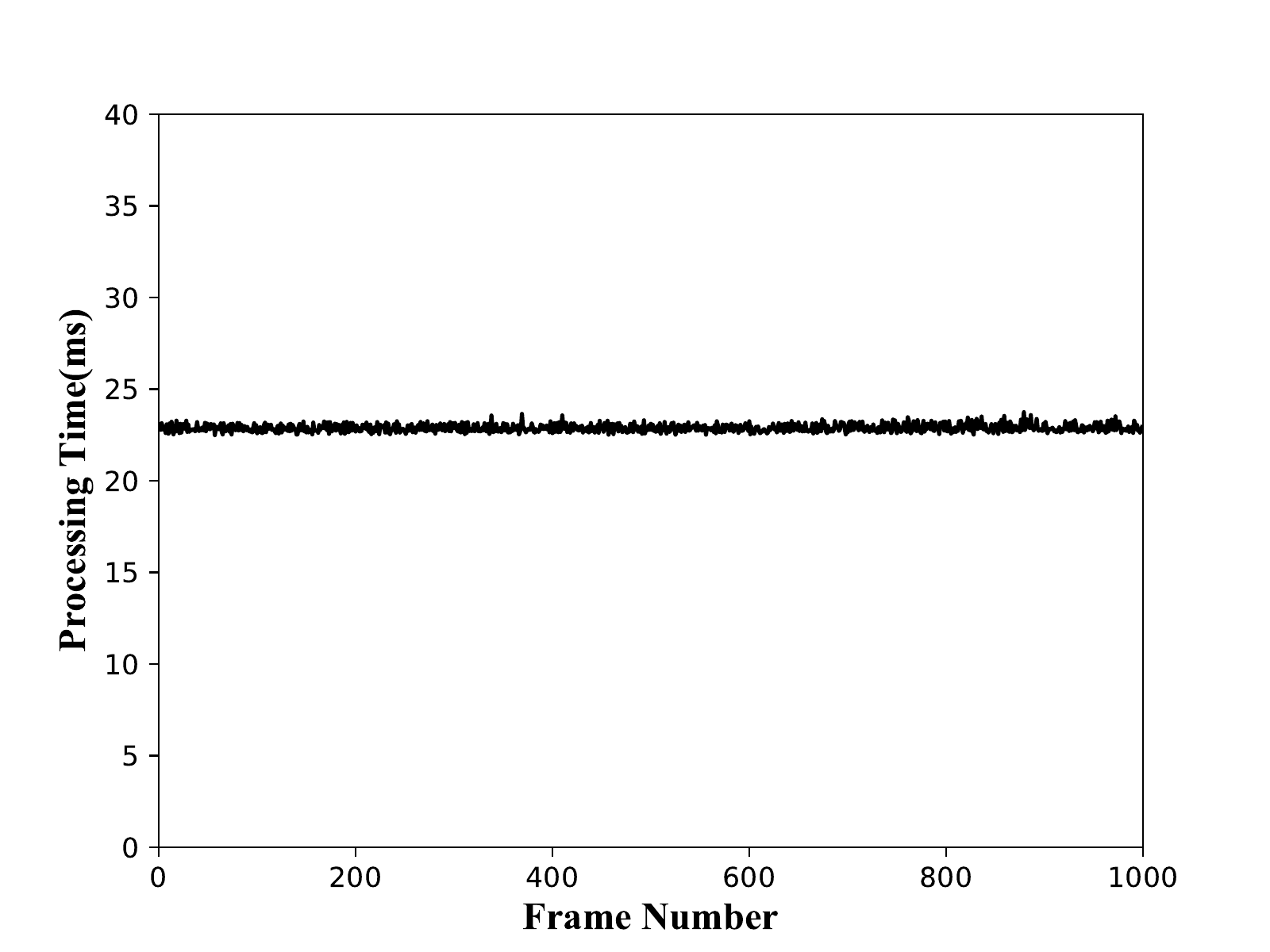}
  \caption{DeepPicar's control loop processing times over 1000 input image frames.}
  \label{fig:control-loop-timing}
\end{figure}

\begin{table}[h]
  \centering
  \begin{tabular} {| c | r | r | r | r |}
    \hline
    \textbf{Operation} & \textbf{Mean} & \textbf{Max} &   \textbf{99pct.} & \textbf{Stdev.} \\ \hline
    Image capture        & 1.61  &  1.81 &  1.75  & 0.05 \\ \hline
    Image pre-processing & 2.77  &  2.90 &  2.87  & 0.04 \\ \hline
    CNN inferencing      & 18.49 & 19.30 & 18.99  & 0.20 \\ \hline
    Total Time           & 22.86 & 23.74 & 23.38  & 0.20 \\ \hline
  \end{tabular}
  \caption{Control loop timing breakdown.}
  \label{tbl:control-loop-breakdown}
\end{table}

Figure~\ref{fig:control-loop-timing} shows the measured control loop
processing times of the DeepPicar over 1000 image frames (one per each
control loop). We omit the first frame's processing time for cache
warmup. Table~\ref{tbl:control-loop-breakdown} shows the time
breakdown of each control loop. Note that all four CPU cores of the
Raspberry Pi 3 were used by the TensorFlow library when performing the
CNN inference operations.

First, as expected, we find that the inference operation
dominates the control loop execution time, accounting for about 81\% of
the execution time.

Second, we find that the measured average
execution time of a single control loop is 22.86 ms, or 43.7 Hz and
the 99 percentile time is 23.38 ms.
This means that the DeepPicar can operate
at up to 40 Hz control frequency in real-time using only the on-board
Raspberry Pi 3 computing platform, as no remote computing resources
were necessary. We consider these results surprising given the complexity
of the deep neural network, and the fact that the inference operation
performed by TensorFlow only utilizes the CPU cores of the Raspberry Pi 3.
In comparison, NVIDIA's DAVE-2 system, which has the exact same neural
network architecture, reportedly runs at 30 Hz~\cite{Bojarski2016}.
Although we believe it was not
limited by their computing platform (we will experimentally compare
performance differences among multiple embedded computing platforms,
including NVIDIA's Jetson TX2, later in
Section~\ref{sec:comparison}), the fact that a low-cost
Raspberry Pi 3 can achieve comparable real-time control performance is
surprising.

Lastly, we find that the control loop execution timing is highly
predictable and shows very little variance over different input image
frames. This is because the amount of computation needed to perform
a CNN inferencing operation is fixed at the CNN architecture design
time and does not change at runtime over different inputs (i.e.,
different image frames). This predictable timing behavior is a highly
desirable property for real-time systems, making CNN inferencing an
attractive real-time workload.

\subsection{Effect of the Core Count to Inference Timing}

In this experiment, we investigate the scalability of performing
inference operations of DeepPicar's neural network with respect to the
number of cores. As noted earlier, the Raspberry Pi 3 platform has
four Cortex-A53 cores and TensorFlow
provides a programmable mechanism to adjust how many cores are to be
used by the library. Leveraging this feature, we repeat the
experiment in the previous subsection but with varying
numbers of CPU cores---from one to four.

\begin{figure}[h]
  \centering
  \includegraphics[width=.45\textwidth]{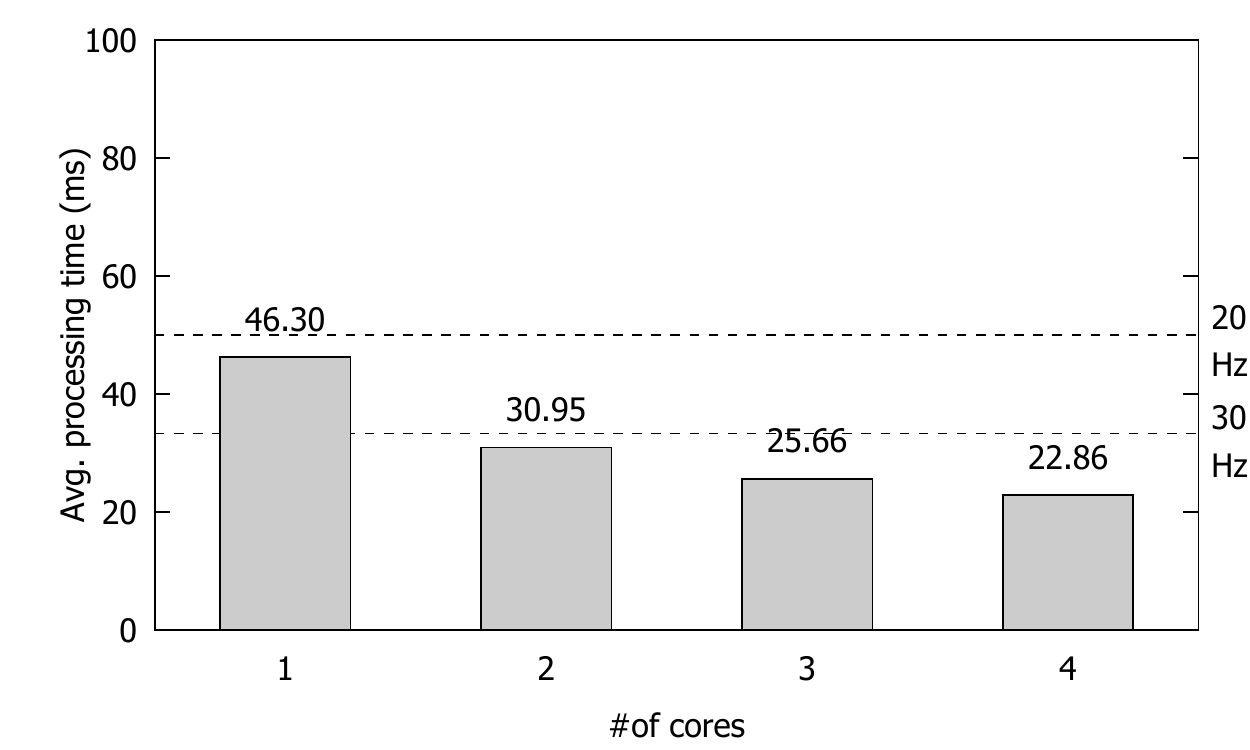}
  \caption{Average control loop execution time vs. \#of CPU
    cores.}
  \label{fig:perf-vs-corecnt}
\end{figure}

Figure~\ref{fig:perf-vs-corecnt} shows the average execution time of
the control loop as we vary the number of cores used by
TensorFlow. As expected, as we assign more cores, the average execution
time decreases---from 46.30 ms on a single core to 22.86 ms on four
cores (over a 100\% improvement). However, the improvement is far from an ideal
linear scaling. In particular, from 3 cores to 4 cores, the
improvement is a mere 2.80 ms, or 12\%. In short, we find that the
scalability of DeepPicar's deep neural network is not ideal.

As noted in~\cite{NVIDIA2015}, CNN inferencing is inherently more
difficult to parallelize than training because the easiest
parallelization option, batching (i.e., processing multiple images in
parallel), is not available or is limited. Specifically, in DeepPicar,
only one image frame, obtained from the camera, can be processed at a
time. Thus, more fine-grained algorithmic parallelization is needed to
improve inference performance~\cite{NVIDIA2015}, which generally does
not scale well.

On the other hand, the limited scalability opens up the possibility of
consolidating multiple different tasks or different neural network
models rather than allocating all cores for a single neural network
model.
For example, it is conceivable to use four cameras and four different
neural network models, each of which is trained separately for a
different purpose and executed on a single dedicated core.
Assuming we use the same network
architecture for all models, then one might expect to achieve up to
20 Hz using one core (given that 1 core can deliver 46 ms average
execution time).
In the next experiment, we investigate the
feasibility of such a scenario.

\subsection{Effect of Co-scheduling Multiple CNN Models}

In this experiment, we launch multiple instances of DeepPicar's CNN
model at the same time and measure its impact on their inference
timings. In other words, we are interested in how shared resource
contention affects inference timing. For this, we create four different
neural network models, that have the same network architecture, and
run them simultaneously.
%% These instances are not identical
%% copies of the same model, but are instead copies of different models
%% (but of the same network architecture).
%% This is done to ensure that no L2 cache memory is shared between the
%% models as they are running.

\begin{figure}[h]
  \centering
  \includegraphics[width=.45\textwidth]{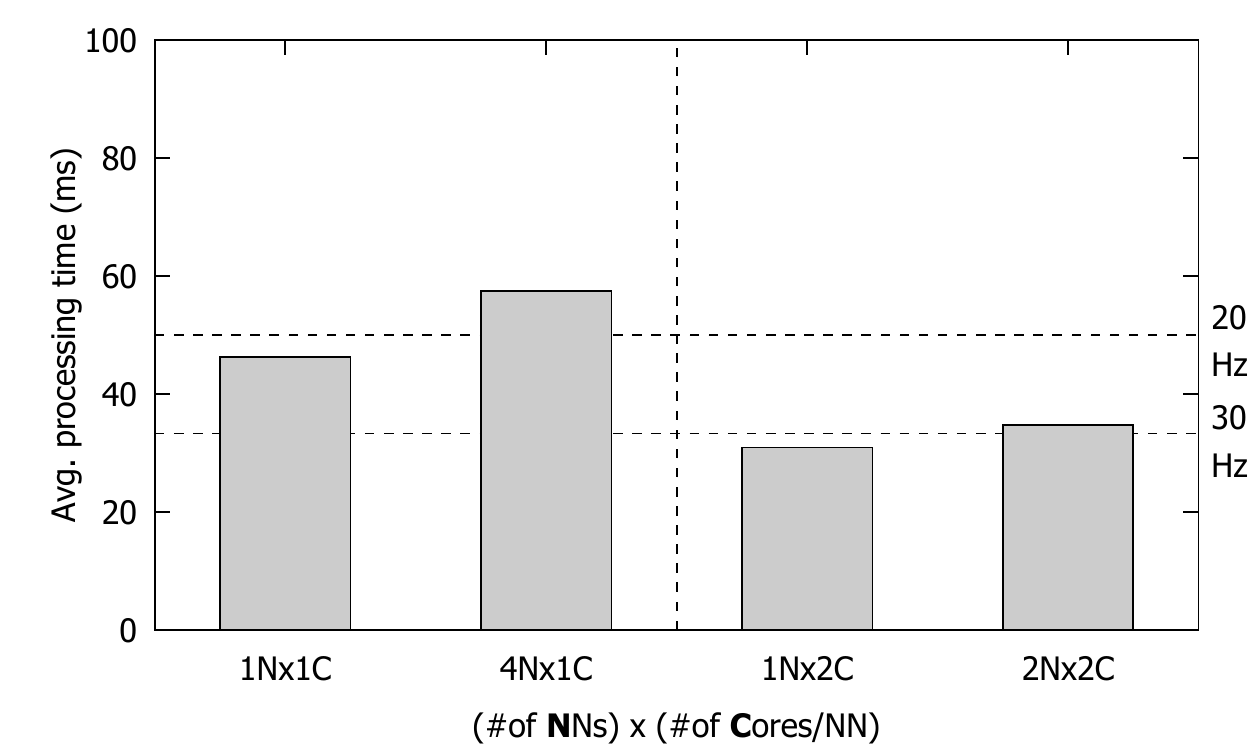}
  \caption{Timing impact of co-scheduling multiple CNNs. 1Nx1C: one CNN
    model using one core; 4Nx1C: four CNN models each using one core;
    1Nx2C: one CNN model using two cores; 2Nx2C: two CNN models each
    using two cores.}
  \label{fig:perf-vs-modelcnt}
\end{figure}

Figure~\ref{fig:perf-vs-modelcnt} shows the results. In the figure, the
X-axis shows the system configuration: \#of CNN models x \#of CPU
cores/CNN. For example, `4Nx1C' means running four CNN models each of
which is assigned to run on one CPU core, whereas `2Nx2C' means running
two CNN models, each of which is assigned to run on two CPU
cores. The Y-axis shows the average inference timing.
The two bars on the left show the impact of co-scheduling four CNN
models. Compared to executing a single CNN model on one CPU core
(1Nx1C), when four CNN models are co-scheduled (4Nx1C), each model
suffers an average inference time increase of approximately 11 ms,
or 24\%. On the other hand, when two CNN models, each using two CPU
cores, are co-scheduled (2Nx2C), the average inference timing is increased by
about 4 ms, or 13\%, compared to the baseline of running one model
using two CPU cores (1Nx2C).

These increases in inference times in the co-scheduled scenarios are
expected because co-scheduled tasks on a
multicore platform interfere with each other due to contention in the
shared hardware resources, such as the shared
cache and DRAM~\cite{Gracioli2015,Yun2013}.

\subsection{Effect of Co-scheduling Synthetic Memory Intensive
  Tasks}\label{sec:eval-memhog}

In this experiment, we investigate the \emph{worst-case} impact of shared
resource contention on DeepPicar's CNN inference timing using
a synthetic memory benchmark. Specifically, we use the \emph{Bandwidth}
benchmark from the IsolBench suite~\cite{Valsan2016}, which
sequentially reads or writes a big array; we henceforth refer to BwRead
as Bandwidth with read accesses and BwWrite as the one with write
accesses. The experiment setup is as follows: We run a single CNN
model on one core, and co-schedule an increasing number of the
Bandwidth benchmark instances on the other cores. We repeat the
experiement first with BwRead and next with BwWrite.

%% ~\footnote{\texttt{\$ bandwdith -a read  -m 16384 -t 1000}}
%% ~\footnote{\texttt{\$ bandwdith -a write -m 16384 -t 1000}}.

\begin{figure}[h]
  \centering
  \includegraphics[width=.45\textwidth]{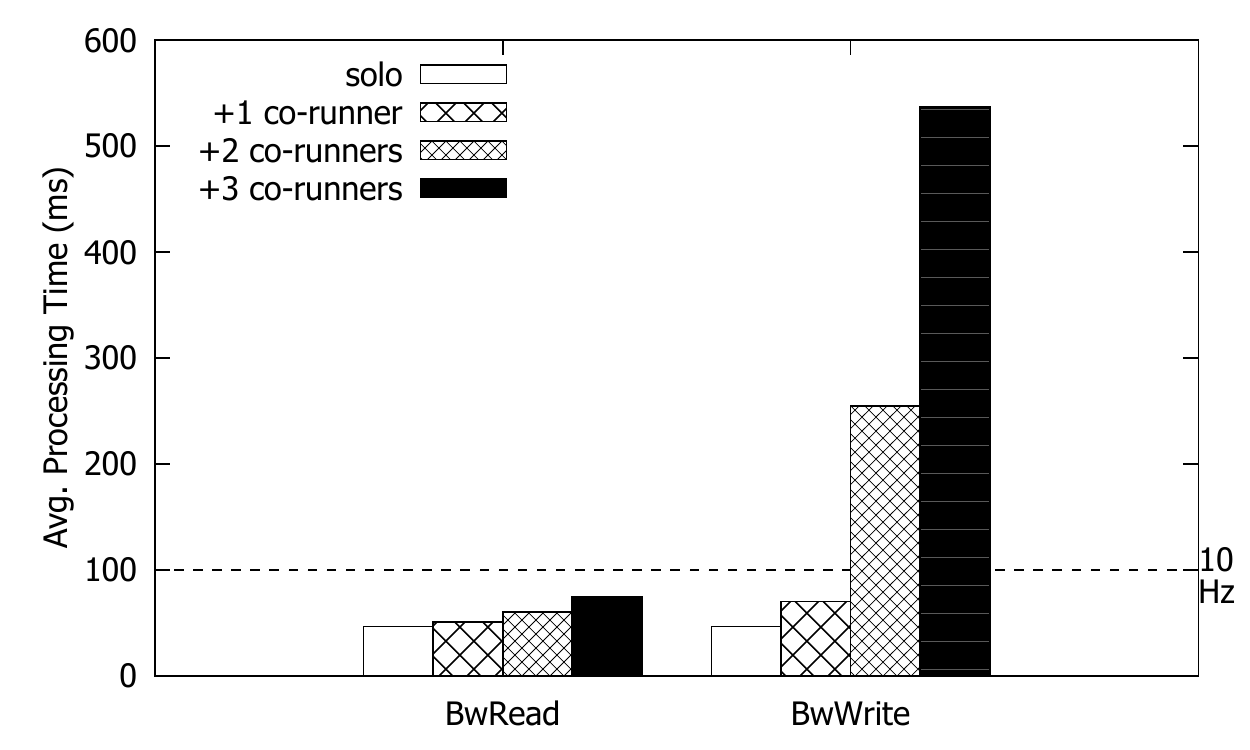}
  \caption{Average processing time vs. the number of memory
    intensive co-runners introduced.}
  \label{fig:perf_vs_bandwidth}
\end{figure}

%% \begin{figure}[h]
%%   \centering
%%   \includegraphics[width=.45\textwidth]{figs/perf_vs_bandwidth_mem}
%%   \caption{Effect of memory performance hogs on the shared resources.
%%     The DNN model uses Core 0 and memory-hog co-runners
%%     use the rest of the cores.}
%%   \label{fig:perf_vs_bandwidth_mem}
%% \end{figure}

Figure~\ref{fig:perf_vs_bandwidth} shows the results. Note
that BwWrite co-runners cause significantly higher execution time
increases---up to 11.6X---on CNN inferencing while BwRead co-runners
cause relatively much smaller time increases. While execution time
increases are expected, the degree to which it is seen in the worst-case
is quite surprising---\emph{if the CNN controller was driving an actual
car, it would result in a car crash!}

Given the importance of predictable timing for real-time control, such
as our CNN based control task, we wanted to: (1) understand the main
source of the timing and (2) evaluate existing isolation methods to
avoid this kind of timing interference. Specifically, shared cache
space and DRAM bandwidth are the two most well-known sources of contention
in multicore systems. Thus, in the following sections, we investigate whether
and to what extent they influence the observed timing interference and
the effectiveness of existing mitigation methods.

\subsection{Effect of Cache Partitioning}

%% \begin{figure}[h]
%%   \centering
%%   \includegraphics[width=.7\textwidth]{figs/palloc_multicore_l2missrate}
%%   \caption{L2 miss rate impact of limiting the amount of L2 cache space
%%   available to the DNN.}
%%   \label{fig:palloc_multicore_l2missrate}
%% \end{figure}

%% \begin{figure}[h]
%%   \centering
%%   \includegraphics[width=.7\textwidth]{figs/palloc_bandwidth_modelbus}
%%   \caption{Model bus access impact of co-scheduling memory intensive co-runners 
%% when cache partitioning is enabled.}
%%   \label{fig:palloc_bandwidth_modelbus}
%% \end{figure}

%% \begin{figure}[h]
%%   \centering
%%   \includegraphics[width=.7\textwidth]{figs/palloc_bandwidth_bus}
%%   \caption{Total bus access impact of co-scheduling memory intensive co-runners 
%% when cache partitioning is enabled.}
%%   \label{fig:palloc_bandwidth_bus}
%% \end{figure}

%% \begin{figure}[h]
%%   \centering
%%   \includegraphics[width=.7\textwidth]{figs/palloc_bandwidth_l2missrate}
%%   \caption{L2 miss rate impact of co-scheduling memory intensive co-
%% runners when cache partitioning is enabled.}
%%   \label{fig:palloc_bandwidth_l2missrate}
%% \end{figure}

Cache partitioning is a well-known technique to improve isolation in
a multicore system by giving a dedicated cache space to each individaul 
task or core. In this experiment, we use PALLOC~\cite{yun2014rtas}, a
page-coloring based kernel-level memory allocator for Linux.
Page coloring is an OS technique that controls the physical addresses
of memory pages. By allocating pages over non-overlapping cache sets,
the OS can effectively partition the cache.
Using PALLOC, we investigate the effect of cache partitioning on
protecting DeepPicar's CNN based controller.

\begin{figure} [h]
  \centering
  \includegraphics[width=.5\textwidth]{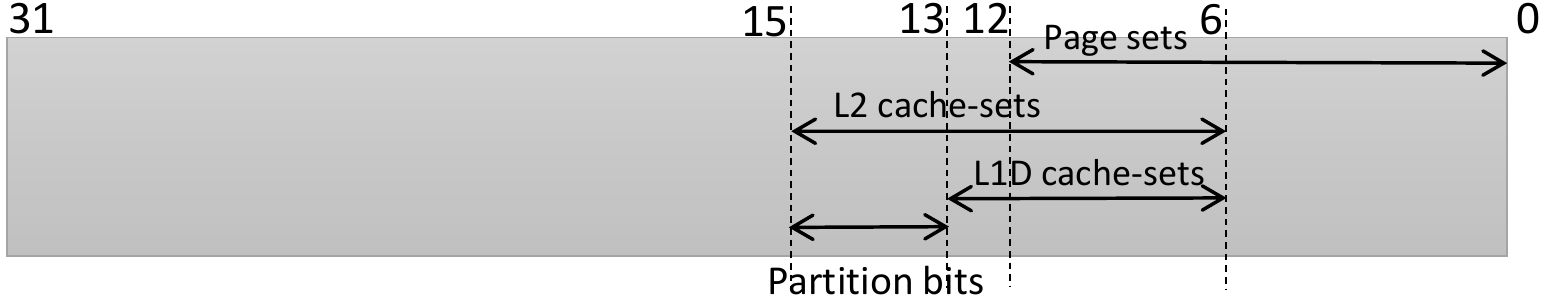}
  \caption{Physical address mapping of L1/L2 caches of Broadcom
    BCM2837 processor in Raspberry Pi 3.}
  \label{fig:cache-mapping}
\end{figure}

Figure~\ref{fig:cache-mapping} shows the physical address
mapping of the Raspberry Pi 3's BCM2837 processor, which has 32K private
L1 I\&D (4way) caches and a shared 512KB L2 (16 way) cache. In order
to avoid partitioning the private L1 caches, we use bits 13 and 14 for
coloring, which results in 4 usable page colors.

%% determine if the DNN is mostly affected by the DRAM 
%% controller, we partition the L2 cache of the Pi3 and reperform the 
%% same experiments to see if there are any noticable changes. For the 
%% partition, we employ PALLOC~\cite{yun2014rtas}, a color-based page 
%% allocator that works on the kernel level.

%% For the bit mask, we select 
%% bits 12, 13 and 14 as they can be used to access the L2 cache,
%% as can be seen by Figure \ref{fig:cache-mapping}.
%% This results in 
%% 2\textsuperscript{3} = 8 colors, which we then assign to the Pi3's 
%% physical cores such that each core has two unique colors (colors 0 
%% and 1 are assigned to core 0, colors 2 and 3 are assigned to core 1, 
%% etc.). In the case of bit 12, since we use 2 colors for each partition,
%% the L1 Data cache is not partitioned in our experiments.

In the first experiment, we investigate the cache space sensitivity of
the DeepPicar's CNN-based control loop. Using PALLOC, we create 4
different cgroups which are configured to use 4, 3,
2, and 1 colors (100\%, 75\%, 50\% and 25\% of the L2 cache
space, respectively). We then execute the CNN control loop (inference)
on one core using a different cgroup cache partition, one at a time,
and measure the average processing time.

\begin{figure}[h]
  \centering
  \includegraphics[width=.45\textwidth]{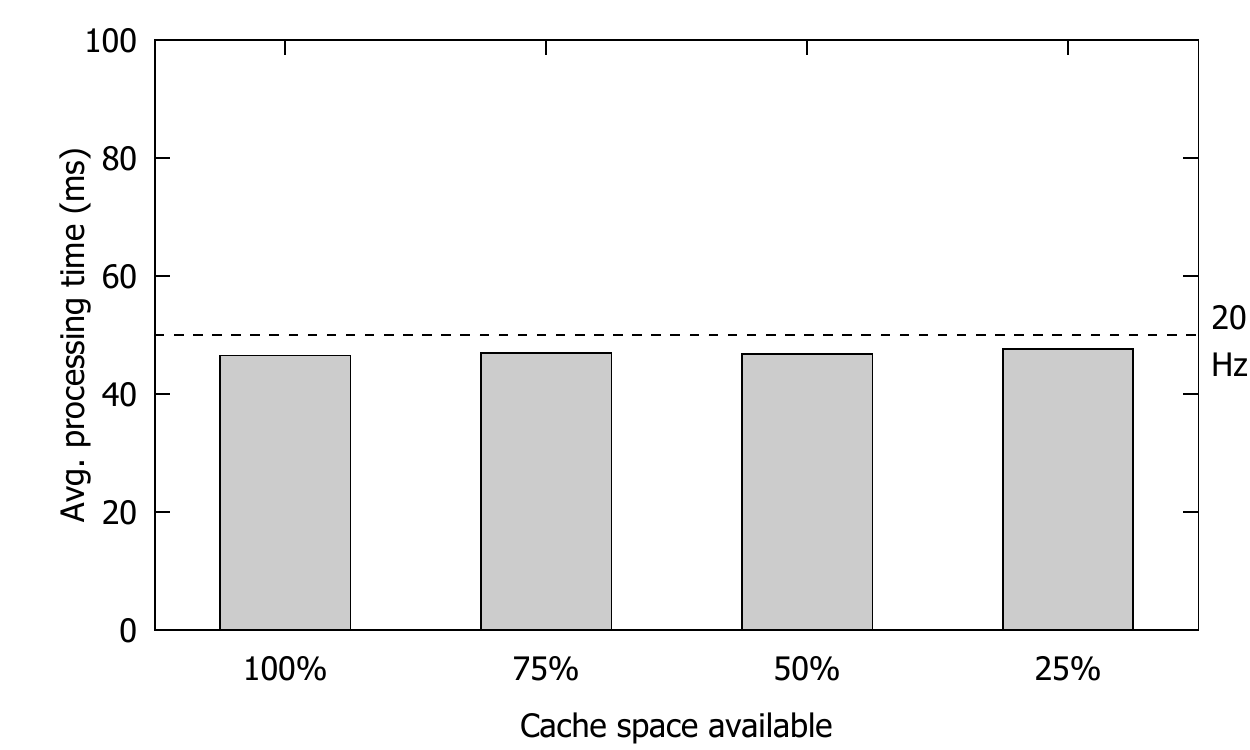}
  \caption{Cache space sensitivity of the CNN controller.}
  \label{fig:palloc_multicore}
\end{figure}

Figure~\ref{fig:palloc_multicore} shows the results. As can be seen,
the CNN inference timing hardly changes at all regardless of the
size of the allocated L2 cache space. In other words, we find that
the CNN workload is largely insensitive to L2 cache space.

The next experiment further validates this finding. In this
experiment, we repeat the experiment in
Section~\ref{sec:eval-memhog}---i.e., co-scheduling the CNN model and
three Bandwidth (BwRead or BwWrite) instances---but this time we
ensure that each task is given equal amounts of L2 cache space by
assigning one color to each task's cache partition.

\begin{figure}[h]
  \centering
  \includegraphics[width=.45\textwidth]{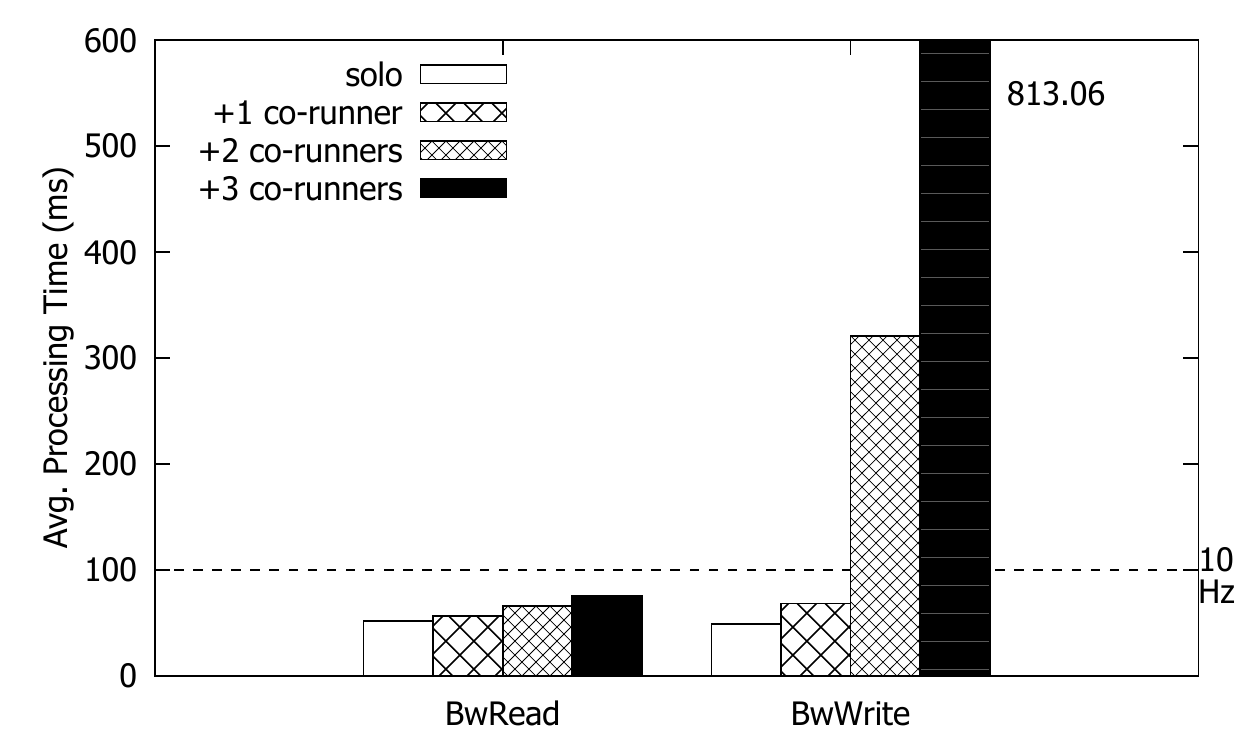}
  \caption{Average processing time vs. the number of memory
intensive co-runners; Each core (task) is given an equal-sized
dedicated cache partition.}
  \label{fig:palloc_bandwidth_exectime}
\end{figure}

Figure \ref{fig:palloc_bandwidth_exectime} shows the
results. Compared to Figure~\ref{fig:perf_vs_bandwidth} where no cache
partioning is applied, assigning a dedicated L2 cache parititon to
each core does not provide significant isolation benefits. For BwRead
co-runners, cache partitioning slightly improves isolation, but for
BwWrite co-runners, cache partitioning causes worse worst-case
slowdown.

In summary, we find that the CNN inferencing workload is not sensitive
to cache space and that cache partitioning is not effective in
providing timing isolation for our CNN workload.

\subsection{Effect of Memory Bandwidth Throttling}

In this subsection, we examine the CNN workload's memory bandwidth
sensitivity and the effect of memory bandwidth throttling in providing
isolation. For the experiments, we use MemGuard\cite{Yun2013}, a Linux
kernel module that can limit the amount of memory bandwidth each core
receives. MemGuard
operates periodically, at a 1 ms interval, and uses hardware
performance counters to throttle cores if they exceed their given
bandwidth budgets within each regulation period (i.e., 1 ms), by
scheduling high-priority idle kernel threads until the next period begins.

In the first experiment, we measure the performance of the CNN model
on a single core, first w/o using MemGuard and then w/
using MemGuard while varying the core's bandwidth throttling parameter
from 500 MB/s down to 100 MB/s.

\begin{figure}[h]
  \centering
  \includegraphics[width=.45\textwidth]{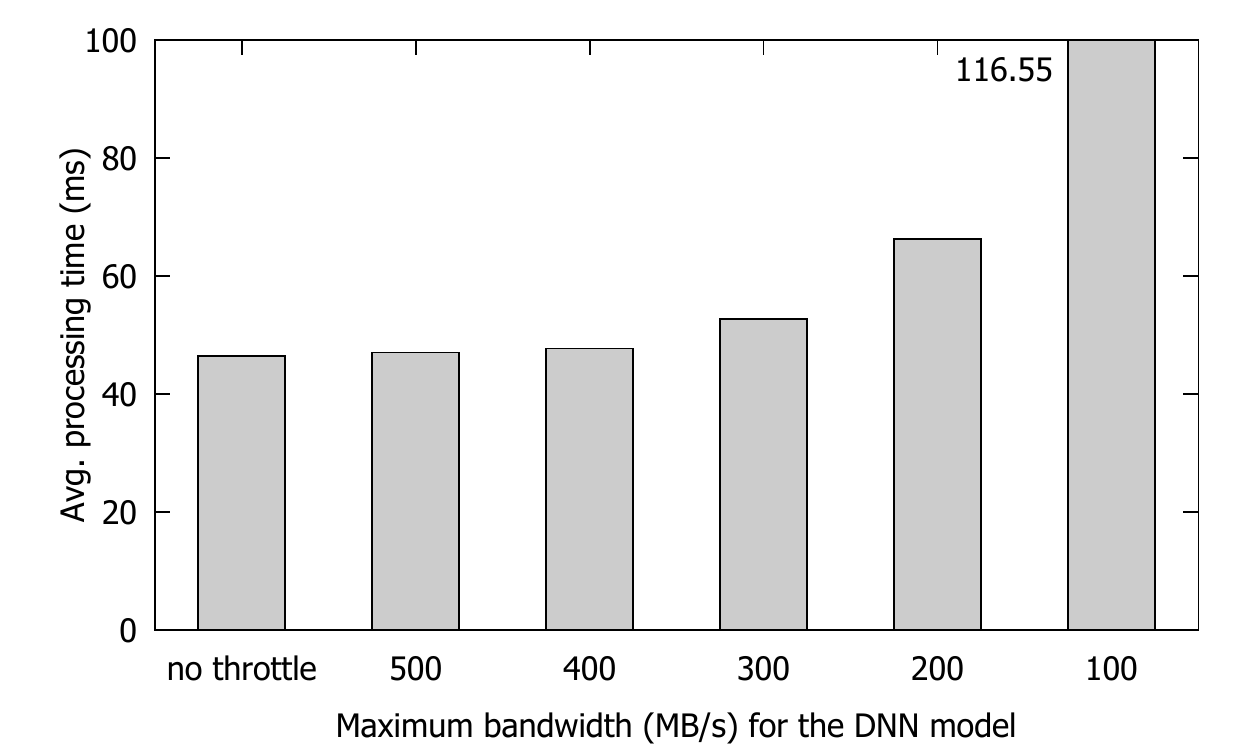}
  \caption{Memory bandwidth sensitivity of the CNN control loop.}
  \label{fig:memguard_multicore}
\end{figure}

Figure \ref{fig:memguard_multicore} shows the results. When the core
executing the CNN model is throttled at 400 MB/s or more, the performance
of the model is largely the same as the non-throttled case. However, as
we decrease the assigned memory bandwidth below 300 MB/s, we start to
observe noticeable decreases in the model's performance. In other
words, the CNN model is sensitive to memory bandwidth and it
requires 400 MB/s or more bandwidth to ensure ideal performance.

In the next experiment, we repeat the experiment in
Section~\ref{sec:eval-memhog}---i.e., co-scheduling memory intensive
synthetic tasks---but this time we
throttle the cores of the co-runners using MemGuard and vary their
memory bandwidth budgets to see their impact on the CNN model's
performance.

\begin{figure}[h]
  \centering
  \includegraphics[width=.45\textwidth]{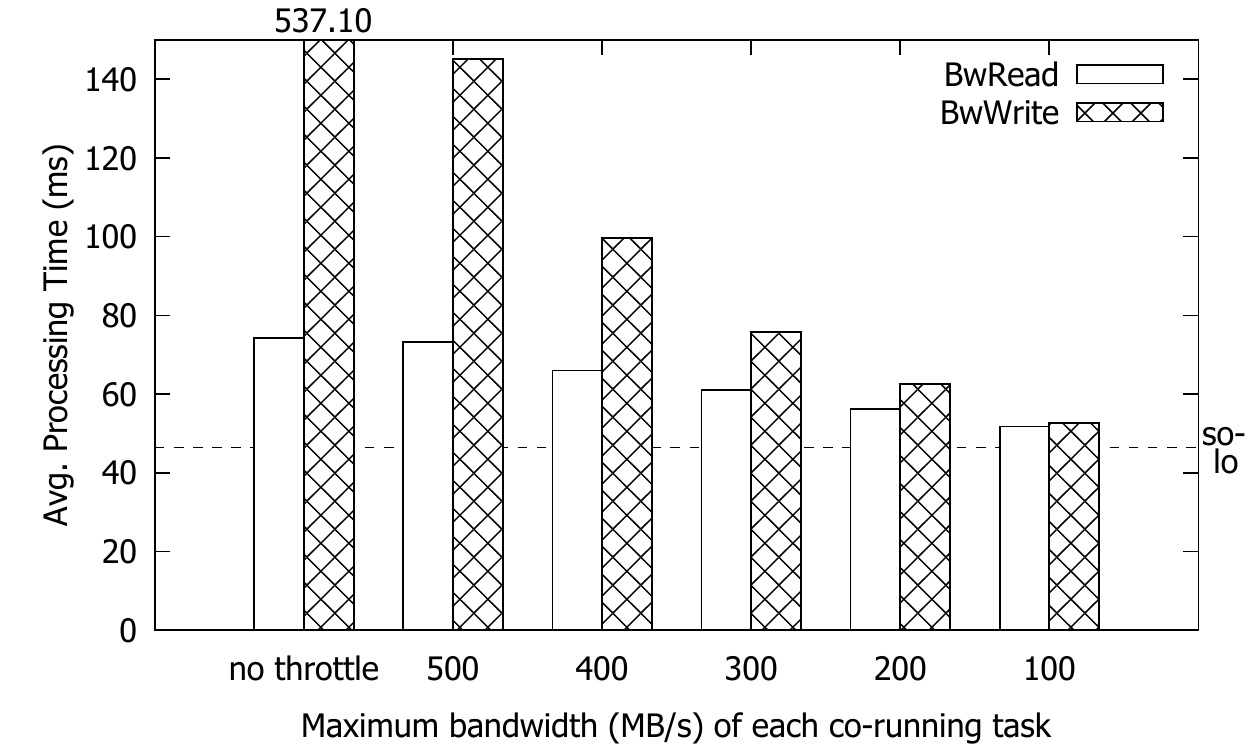}
  \caption{Effect of throttling three memory intensive co-runners.}
  \label{fig:memguard_bandwidth}
\end{figure}

Figures \ref{fig:memguard_bandwidth} shows the results.
As can clearly be seen in the figure, limiting the co-runners's memory
bandwidth is effective in protecting the CNN model's performance for
BwRead and BwWrite co-runners. The benefits are especially more
pronounced in case of BwWrite co-runners as, when we throttle them more, the
CNN's performance quickly improves.

%% Note that, in case of BwWrite co-runners, we assign half the
%% bandwidth budget because MemGuard only accounts for L2 refills
%% but not write-backs, which effectively allows twice the allocated
%% memory bandwidth budget.

In summary, we find that the CNN inferencing workload is sensitive to
memory bandwidth and that memory bandwidth throttling is effective in
improving the performance isolation of the CNN workload.

\section{Embedded Computing Platform Comparison}\label{sec:comparison}

%% \begin{figure}[h]
%%   \centering
%%   \includegraphics[width=.45\textwidth]{figs/a53_vs_a7}
%%   \caption{DNN model performance on different in-order
%% ARM Cortex cores. \fixme{this may be better in the plaform comparison section}.}
%%   \label{fig:a53_vs_a7}
%% \end{figure}

\begin{table*}[h]
  \centering
  \begin{adjustbox}{width=1\textwidth}
  \begin{tabular}{|c|c|c|c|}
    \hline
    Item    & Raspberry Pi 3 (B)   & Intel UP                  & NVIDIA Jetson TX2\\
    \hline
            & BCM2837              & X5-Z8350 (Cherry Trail)   & Tegra X2 \\
    CPU     & 4x Cortex-A53@1.2GHz/512KB L2  &
              4x Atom@1.92GHz/2MB L2 &
              4x Cortex-A57@2.0GHz/2MB L2 \\
            &              &              & 2x Denver@2.0GHz/2MB L2 (not used)  \\
    \hline
    GPU     &  VideoCore IV (not used)    &
               Intel HD 400 Graphics (not used) &
               Pascal 256 CUDA cores   \\
    \hline
    Memory  & 1GB LPDDR2 (Peak BW: 8.5GB/s)  &  2GB DDR3L (Peak BW: 12.8GB/s)    & 8GB LPDDR4  (Peak BW: 59.7GB/s)            \\
    %% \hline
    %%     Peak Memory Bandwidth & 8.5 GB/s & 12.8 GB/s & 59.7 GB/s \\
	%% \hline
	%% Measured Bandwidth & 2127.94 MB/s & 3951.94 MB/s & 4447.90 MB/s \\
	\hline
	Cost  & \$35 & \$100 & \$600 \\
	\hline
  \end{tabular}
  \end{adjustbox}
  \caption{Compared embedded computing platforms}
  \label{tbl:platforms}
\end{table*}

In this section, we compare three computing platforms---the Raspberry
Pi 3, the Intel UP~\cite{intelup} and the NVIDIA Jetson
TX2~\cite{nvidiajetson}---from the point of view of supporting
end-to-end deep learning based autonomous vehicles. 
Table~\ref{tbl:platforms} shows the architectural features of the three
platforms~\footnote{The GPU of Intel UP and the two Denver cores in the
  Tegra TX2 are not used in evaluation due to TensorFlow issues.}.
  
Our basic approach is to use the same DeepPicar software, and repeat
the experiments in Section~\ref{sec:evaluation} on each hardware
platform and compare the results. 
%We do not, however, repeat the cache
%partitioned experiments done on the Pi 3, as we did not believe our 
%findings warranted further experimentation on additional platforms. 
For the Jetson TX2, we have two different system configurations,
which differ in whether TensorFlow is configured to use its GPU or
only the CPU cores. Thus, a total of four system configurations are
compared.

\begin{figure}[h]
  \centering
  \includegraphics[width=.45\textwidth]{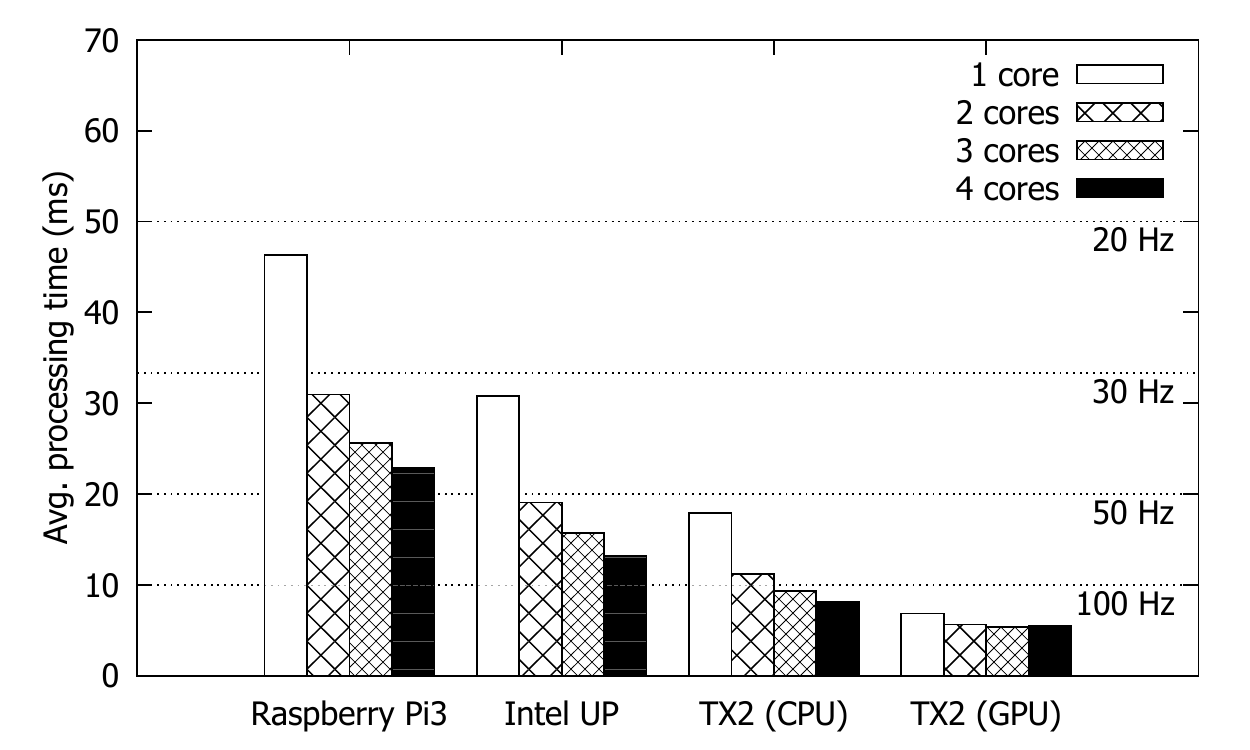}
  \caption{Average control loop execution time.} 
  \label{fig:sys_core}
\end{figure}

Figure~\ref{fig:sys_core} shows the average control loop completion
timing of the four system configurations we tested as a function of
the number of CPU cores used. (cf. Figure~\ref{fig:perf-vs-corecnt})
Both the Intel UP and Jetson TX2 exhibit better performance than
Raspberry Pi 3. When all four CPU cores are used, the Intel UP is
1.33X faster than Pi 3, while the TX2 (CPU) and TX2 (GPU) are 2.79X and
4.16X times faster that the Pi 3, respectively.
Thus, they all satisfy 33.3 ms WCET by a clear margin, and, in the
case of the TX2, 50 Hz or even 100 Hz real-time control is feasible
with the help of its GPU.
Another observation is that the CNN task's performance on TX2 (GPU)
does not change much as we increase the number of cores. This
is because most of the neural network computation is done by the GPU.

\begin{figure}[h]
  \centering
  \includegraphics[width=.45\textwidth]{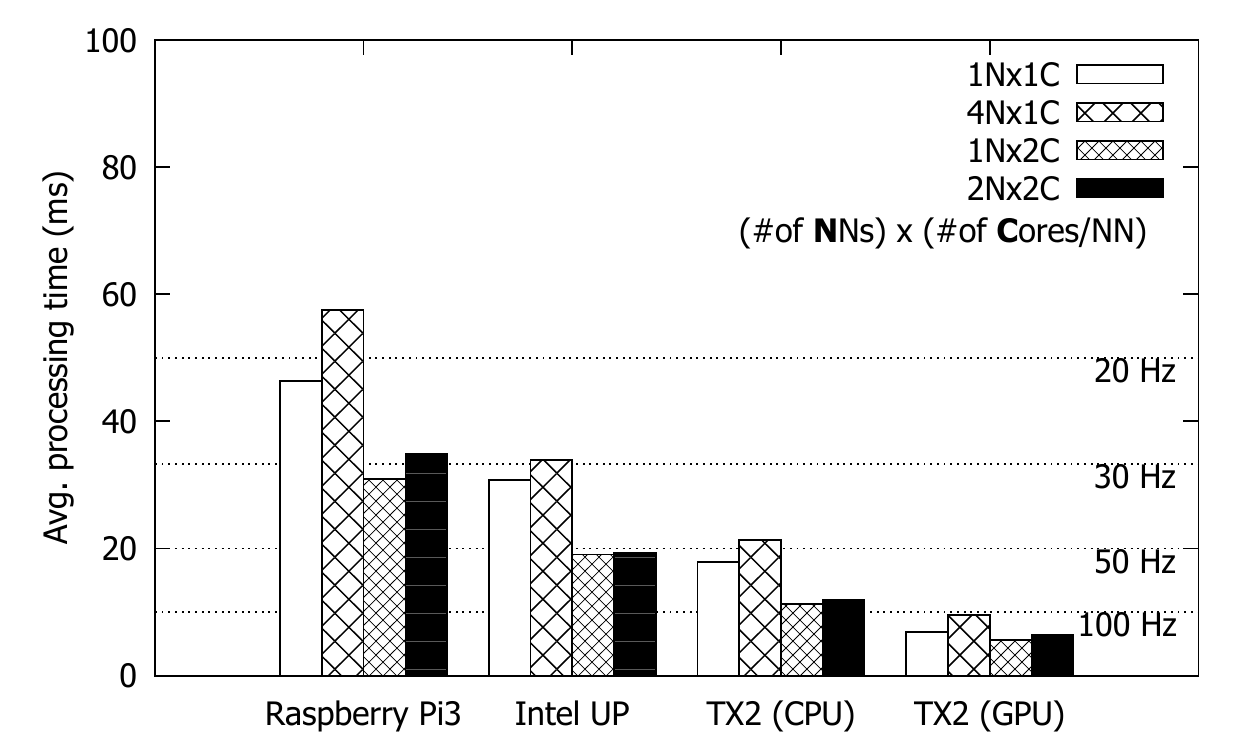}
  \caption{Timing impact of co-scheduling multiple CNNs on different
    embedded multicore platforms. %% 1Nx1C: one DNN
    %% model using one core; 4Nx1C: four DNN models each using one core;
    %% 1Nx2C: one DNN model using two cores; 2Nx2C: two DNN models each
    %% using two cores.
  }
  \label{fig:sys_model}
\end{figure}

Figure~\ref{fig:sys_model} shows the results of the
multi-model co-scheduling experiment
(cf. Figure~\ref{fig:perf-vs-modelcnt}). Once again, they can comfortably
satisfy 30 Hz real-time performance for all of the co-scheduled CNN control
loops, and in the case of the TX2 (GPU), even 100 Hz real-time control
is feasible in all co-scheduling setups.
Given that the GPU must be shared among the co-scheduled CNN
models, the results suggest that the TX2's GPU has sufficient capacity to
accomodate multiple instances of the CNN models we tested.

\begin{figure}[h]
  \centering
  \begin{subfigure}{0.45\textwidth}
    \includegraphics[width=\textwidth]{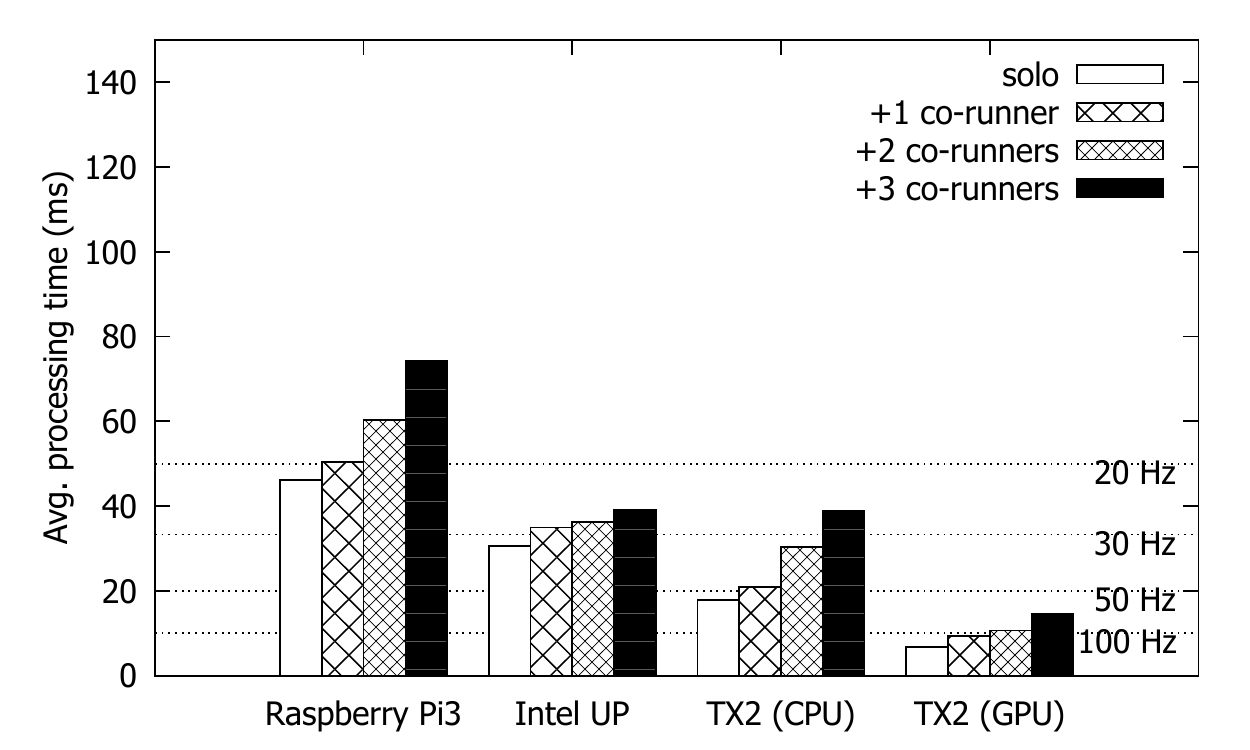}
    \caption{BwRead}
    \label{fig:sys_bench_read}
  \end{subfigure}
  \hfill
  \begin{subfigure}{0.45\textwidth}
    \includegraphics[width=\textwidth]{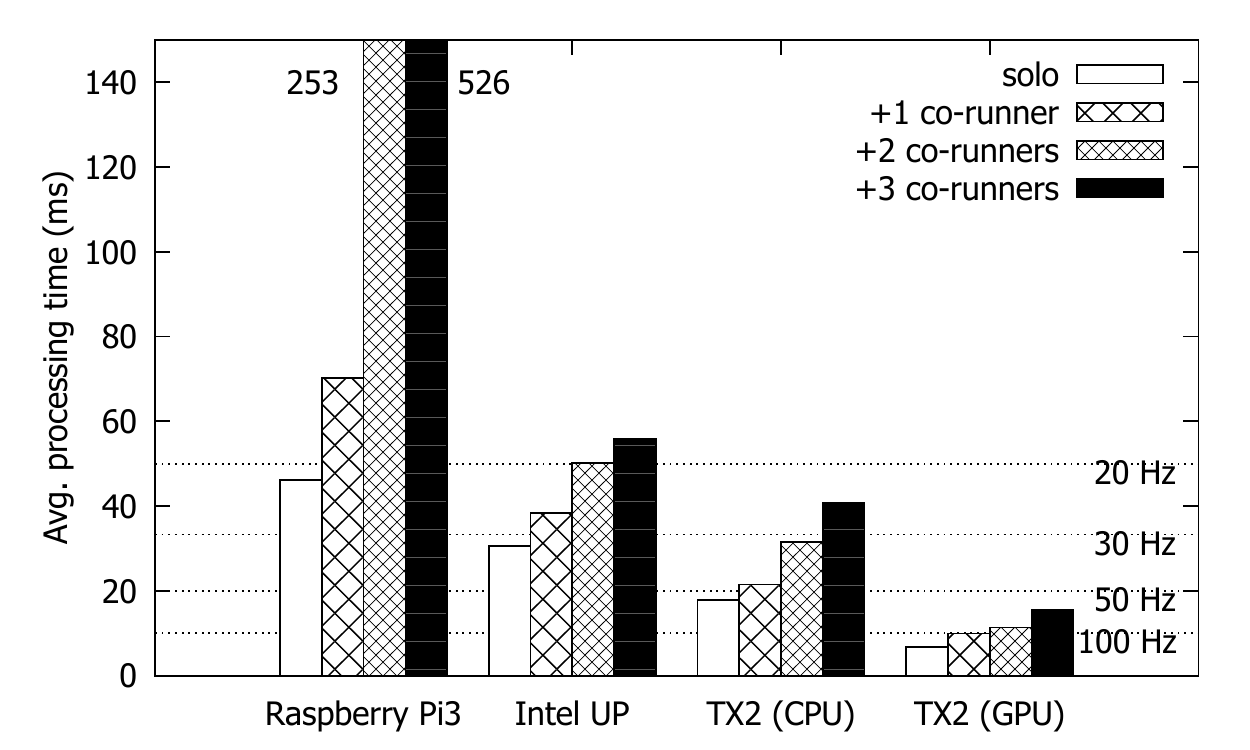}
    \caption{BwWrite}
    \label{fig:sys_bench_write}    
  \end{subfigure}
  \caption{Average processing time vs. the number of memory
    intensive co-runners.}
  \label{fig:sys_bench}    
\end{figure} 

Figure~\ref{fig:sys_bench} shows the results of the synthetic memory
intensive task co-scheduling experiments
(cf. Figure~\ref{fig:perf_vs_bandwidth}).
For read co-runners (BwRead), the performance of all platforms
gradually decreased as additional BwRead instances were introduced: up
to 1.6X for the Pi 3, up to 1.3X for the Intel UP, and up to 1.6X and 2.2X for
the TX2 (CPU) and TX2(GPU), respectively.
For write co-runners (BwWrite), however, we observe generally more
noticieable execution time increases. As we discussed
earlier in Section~\ref{sec:evaluation}, the Pi 3 suffers up to 11.6X
execution time increase, while the Intel UP and Jetson TX2 suffer less
dramatic but still significant execution time increases.

%% Compared to their respective solo timings  
%% (i.e., the model runs on a single core in isolation), Intel UP suffers up to
%% 2.3X execution time increase; TX2 (CPU) and TX2 (GPU) both suffer up to
%% 2.3X increases. This is somewhat surprising
%% because the Raspberry Pi 3's cores are in-order architecture based while
%% the cores in the Intel Up and NVIDIA TX2 are out-of-order architecture
%% based, and that the memory intensive tasks on out-of-order cores can
%% generate more memory traffic.
%% We believe that this is because the
%% memory subsystems in the Intel UP and TX2 platforms provide higher
%% performance than the memory subsystem of the Pi 3 as suggested from
%% the measured memory bandwidth results in Table~\ref{tbl:platforms}
%% ('Measured Bandwidth').

Another interesting observation is that the TX2 (GPU) also suffers
considerable execution time increase (2.3X) despite the fact that the
co-scheduled synthetic tasks do not utilize the GPU (i.e., the CNN
model has dedicated access to the GPU.) This is, however, a 
known characteristic of integrated CPU-GPU architecture based
platforms in which both the CPU and GPU share the same memory
subsystem~\cite{Ali2017} and therefore can suffer bandwidth contention
as we observe in this case.

In summary, we find that todays embedded computing platforms, even as
inexpensive as a Raspberry Pi 3, are powerful enough to support
CNN based real-time control applications. Furthermore, availability of
CPU cores and a GPU on these platforms allows consolidating multiple CNN
workloads. However, shared resource contention among these diverse
computing resources remains an important issue that must be understood
and controlled, especially for safety-critical applications.

\section{Related Work}\label{sec:related}
% discusstion
% - model is still quite small. cam we use bigger CNNs?
%% \fixme{What we want to say:
%%   1) We showed DNN is feasible on today's embedded computing
%%   platforms.
%%   2) But, the CNN model is rather smaller-kind with respect to today's
%%   state-of-the-art CNNs, which are much bigger.
%%   3) disscuss google's object detection api
%%   paper~\cite{huang2017speed}.
%%   4) Applying these much more complex DNNs (such as google's detection
%%   models) may be challenging for embedded computing
%%   platforms.
%%   5) Two general approaches: (1) make it faster (computing workload
%%   remains the same) (2) reduce the workload itself (model compression
%%   or reduced precision etc.)}

%% {\bf DNN optimization for embedded systems.}

There are several relatively inexpensive RC-car based autonomous car
testbeds. MIT's RaceCar~\cite{shin2017project} and UPenn's
F1$/$10~\cite{upennf1tenth} are both based on a Traxxas 1/10 scale RC
car and a NVIDIA Jetson multicore computing platform, which
is equipped with many sophisticated sensor packages, such as a lidar.
However, they both cost more than \$3,000, requiring a considerable
investment. DonkeyCar~\cite{donkeycar} is similar to our DeepPicar as
it also uses a Raspberry Pi 3 and a similar CNN for end-to-end
control, although it costs more (about \$200).
%% because it uses a more
%% expensive 1/16 scale RC car than our 1/24 scale one.
The main contribution of our paper is in the detailed analysis of
computational aspects of executing a CNN-based real-time control
workload on diverse embedded computing platforms.

In this paper, we have analyzed real-time performance of a real-world
CNN, which was used in NVIDIA's DAVE-2 self-driving
car~\cite{Bojarski2016}, on a low-cost Raspberry Pi 3 quad-core 
platform and other embedded multicore platforms. It should be noted,
however, that DAVE-2's CNN is relatively small compared
to recent state-of-the-art CNNs, which are increasingly larger and
deeper. For example, the CNN based object detector models evaluated in
Google's recent study~\cite{huang2017speed} have between 3M to 54M
parameters, which are much larger than DAVE-2's CNN.
Using such large CNN models will be challenging on
resource constrainted embedded computing platforms, especially for
real-time applications such as self-driving cars.

%% available DNN's can be much larger compared to the architecture of the
%% DAVE-2. Such DNN's can be seen in Google's Object Detection API
%% The smallest model available in the API is the MobileNet model, which
%% has 3,191,072 parameters and is approximately 12.7 larger than the
%% model used by the DeepPicar. The other models all have parameter sizes
%% of at least 10 million. Due to the considerable differences in the
%% model sizes, it is likely that running them on platforms like the Pi 3,
%% and still get acceptable real-time performance would be difficult,
%% if not infeasible.

%% % related work on efficient DNN representation and processing.
%% While repeated feedforward and backpropagation operations during
%% the training time account for most of the computational cost in
%% deep learning, there is a growing need for an improved efficiency
%% during the inferencing time as well, especially because of the
%% potential of utilizing Deep Neural Networks (DNN) for the real-time
%% pattern recognition tasks in embedded systems, as our paper
%% exemplifies.

While continuing performance improvements in embedded computing
platforms will certainly make processing these complex CNNs faster,
another actively investigated approach is to reduce the required
computational complexity itself.
%% When a DNN is deployed in those implementations with limited
%% resources, such as memory and power, the floating-point operations
%% involved in the large matrix multiplications are a burdensome task.
Many recent advances in network compression have shown promising results
in reducing such computational costs during the feedforward
process. The fundamental assumption in those techniques is that the
CNNs are redundant in their structure and representation. For example,
network pruning can thin out the network and provides a more condensed
topology~\cite{han2015deep}.

Another common compression method is to reduce the
quantization level of the network parameters, so that arithmetic
defined with floating-point operations are replaced with low-bit
fixed-point counterparts. To this end, single bit quantization of the
network parameters or ternary quantization have been recently proposed
~\cite{hwang2014fixed,soudry2014expectation,kim2016bitwise,rastegari2016xnor,hubara2016binarized,beauchamp2006embedded,govindu2004analysis}.
In those networks, the inner product between the
originally real-valued parameter vectors is defined with XNOR followed
by bit counting, so that the network can greatly minimize the
computational cost in the hardware implementations. This drastic
quantization can produce some additional performance loss, but those
new binarized or ternarized systems provide a simple quantization
noise injection mechanism during training so that the additional error
is minimized to an acceptable level.

The XNOR operation and bit counting have been known to be efficient in
hardware implementations. In~\cite{rastegari2016xnor}, it was shown
that the binarized convolution could substitute the expensive
convolutional feedforward operations in a regular CNN, by using only
about 1.5\% of the memory space, while
providing 20 to 60 times faster feedforward. Binary weights were also
able to provide 7 times faster feedforward than a floating-point
network for the hand written digit recognition task as well as 23
times faster matrix multiplication tasks on a
GPU~\cite{hubara2016binarized}.
Moreover, FPGA implementations showed that the XNOR operation is 200
times cheaper than floating-point multiplications with
single precision~\cite{beauchamp2006embedded,govindu2004analysis}.
XNOR-POP is another hardware implementation that reduced
the energy consumption of a CNN by 98.7\%~\cite{jiang2017xnor}.

These research efforts are expected to make complex CNNs accessible
for a wider range of real-time embedded systems. We plan to
investigate the feasibility of these approaches in the context of
DeepPicar so that we can use even more resource constrained
micro-controller class computing platforms in place of the current
Raspberry Pi 3.

%% % related work on testbeds
%% \fixme{maybe we can also discuss MIT racecar, UPENN car, donkey car and other testbeds?
%%   The points of discussion may include: 1) they are expensive. 2)
%%   donkey car is similar to us, but still a bit more expnsive, and more
%%   importantly there is no existing studies which systematically evaluate
%%   real-time performance of using DNN on embedded computing
%%   platforms. Our work provides quantitive real-time performance
%%   evaluation results, using a known DNN used in a real self-driving
%%   car. Our work is fully reproducible at a very low cost, lowring the
%%   barrier to entry in researching self-driving cars, helping the
%%   community blah blah blah.}

%% and the ability of the car to navigate the desired environment. We are
%% more focused on the real-time performance of the DeepPicar, and whether
%% the Raspberry Pi 3 embedded computing platform can properly support autonomous
%% vehicle operations. To the best of our knowledge, no other studies have
%% been done to evaluate the capabilities of the Pi 3 for running DNNs and
%% performing autonomous vehicle operations.

\section{Conclusion}\label{sec:conclusion}
We presented DeepPicar, a low cost autonomous car platform that is
inexpensive to build, but is based on state-of-the-art AI technology:
End-to-end deep learning based real-time control.
Specifically, DeepPicar uses a deep convolutional neural network to
predict steering angles of the car directly from camera input data
in real-time. Importantly, DeepPicar's neural network architecture is
identical to that of NVIDIA's real self-driving car DAVE-2. 

Despite the complexity of the neural network, DeepPicar uses a
low-cost embedded quad-core computer, the Raspberry Pi 3,  as its sole
computing resource.
We systematically analyzed the platform's real-time capability in
supporting the CNN-based real-time control task.
We also evaluated other, more powerful, embedded computing
platforms to better understand achievable real-time performance of
DeepPicar's CNN based control system and the impact of
computing hardware architectures.
We find all tested embedded platforms, including the Pi 3, are capable
of supporting the CNN based real-time control, from 20 Hz up to
100 Hz, depending on the platform.
Futhermore, all platforms were capable of consolidating
multiple CNN models and/or tasks.

However, we also find that shared resource
contention remains an important issue that must be considered to
ensure desired real-time performance on these shared memory based
embedded computing platforms.
Toward this end, we evaluated the impact of shared resource contention
to the CNN workload in diverse consolidated workload setups, and
evaluated the effectivness of state-of-the-art shared resource
isolation mechanisms in protecting performance of the CNN
based real-time control workload.

%% As future work, we plan to improve the prediction accuracy of the
%% system by feeding more data in training.
As future work, we plan to investigate ways to reduce computational
and memory overhead of CNN inferencing and to evaluate the
effectiveness of FPGA and other specialized accelerators.

%% we plan to apply shared DRAM resource management
%% techniques~\cite{Yun2013,yun2014rtas} on the DeepPicar platform and
%%evaluate their impact on the real-time performance of the system. 

%% and
%% upgrading the RC car hardware platform to enable more precise steering
%% angle control

\section*{Acknowledgements} \label{acknowledge}
This research is supported in part by NSF CNS 1815959 and the National Security Agency (NSA) Science of Security Initiative.
The Titan Xp and Jetson TX2 used for this research were donated by the
NVIDIA Corporation.

%-------------------------------------------------------------------------
\bibliographystyle{abbrv}
\bibliography{reference}

\appendix
%% \begin{figure*}[t]
%%   \centering
%%   \includegraphics[width=.9\textwidth]{figs/TrainingLoss}
%%   \caption{Change in loss value throughout training.}
%%   \label{fig:modelloss}
%% \end{figure*}

\subsection{DNN Training and Testing}
We have trained and tested the deep neural network with several
different track conditions, different combinations of input
data, and different hyper parameters. In the following paragraphs, we
describe details on two of the training methods that performed
reasonably well.

In the first method, we trained the neural network model across a set
of 30 completed runs on the track seen in Figure~\ref{fig:track2} by a
human pilot. Half of the runs saw the car driving one way along the
track, while the remaining half were of the car driving in the
opposite direction on the track.
In total, we collected 2,556 frames for training and 2,609
frames for validation.
The weights of the network are initialized using the Xavier
initializer~\cite{Glorot2010}, which is known to perform better
than a random weight assignment.
In each training step, we use a batch
size of 100 frames, which are randomly selected among all the
collected training images, to optimize the network.
We repeat this across 2,000 training steps.
%% When a model was trained
%% with the  aformentioned data, the training loss was 0.0188 and the
%% validation  loss was 0.0132.
%% Figure~\ref{fig:modelloss} shows the change of the loss value over the
%% course of model training.

In the second method, we use the same data and parameters as  above
except that now images are labeled as `curved' and `straight' and we pick an
equal number of images from each category at each training step to
update the model. In other words, we try to remove bias in selecting
images. We find that the car performed better in practice by applying
this approach as the car displayed a greater ability to stay in the
center of the track (on the white tape).
However, we find that there is a discrepency between the training
loss and the validation loss, indicating that the model may suffer
from an overfitting problem, despite its better real-world
performance.

%% as the former was 0.009, while the latter
%% was 0.0869

%% We plan to investigate ways to achieve better prediction accuracy
%% in training the network. We also plan to evaluate different RC
%% car platforms that have different steering mechanisms to see their
%% impact to the control performance.

\subsection{System-level Factors Affecting Real-Time Performance}
In using the Raspberry Pi 3 platform, there are a
few system-level factors, namely power supply and temperature, that
need to be considered to achieve consistent performance.

In all of our experiments on the Raspberry Pi 3, the CPU  is configured
at the maximum clock speed of 1.2 GHz. However, without care, the CPU can
operate at a lower frequency involuntarily.
An important factor is CPU thermal throttling, which can affect CPU
clock speed if the CPU temperature is too high (Pi 3's firmware is
configured to throttle at 85 deg. C).
DNN inferencing is computationally intensive, thus the temperature of
the CPU could rise quickly. This can be especially problematic in
situations where multiple DNN models run simultaneously on the
Pi 3. If the temperature reaches the threshold, the Pi 3's thermal
throttling kicks in and decreases the clock speed down to 600MHz---
half of the maximum 1.2GHz---so that the CPU's temperature stays at a
safe level.
We found that without proper cooling solutions (heatsink or fan),
prolonged use of the system would result in CPU frequency decrease
that may affect evaluation.

Another factor to consider is power supply. The
Pi 3 frequency throttling also kicks in when the power source can not
provide 2A current.
In experiments conducted with a power supply that only provided 1 Amp,
the Pi was unable to sustain a 1.2 GHz clock speed. As a result, it
is necessary, or at least highly recommended, that the power supply
used for the Raspberry Pi 3 be capable of outputting 2 Amps, otherwise
optimal performance isn't guaranteed.

Our initial experiment results suffered from these issues, after which
we always carefully monitored the current operating frequencies of the
CPU cores during the experiments to ensure the correctness and
repeatability of the results.

\end{document}